\documentclass[aps,prd,amsmath
,eqsecnum            
,preprintnumbers
,nofootinbib         
 ,showpacs          
 ,showkeys          
,twocolumn         
]{revtex4}
\usepackage[dvips]{graphicx}
\usepackage{amsmath}
\usepackage{amsfonts}
\usepackage{amssymb}
\usepackage{longtable}   

\allowdisplaybreaks[4]     

\newcommand{\df}{\ {\overset {\rm def} =}\ }
\newcommand{\dr}[2]{\frac {{\rm d} {#1}} {{\rm d} {#2}}}

\newcommand{\dril}[2]{{{\rm d} {#1}} / {{\rm d} {#2}}}

\begin{document}

\title{Blueshifts in the Lema\^{\i}tre -- Tolman models}

\author{Andrzej Krasi\'nski}
\affiliation{N. Copernicus Astronomical Centre, Polish Academy of Sciences, \\
Bartycka 18, 00 716 Warszawa, Poland} \email{akr@camk.edu.pl}

\date {}

\begin{abstract}
In the Lema\^{\i}tre -- Tolman (L--T) models that have nonconstant bang-time
function $t_B(r)$, light emitted close to those points of the Big Bang where
$\dril {t_B} r \neq 0$ is blueshifted at the receiver position. The blueshifted
rays are expected to perturb the temperature of the cosmic microwave background
radiation along the lines of sight of the present central observer. It is shown
here that, in a general L--T model, the $t_B(r)$ can be chosen so that the
blueshift-generating region is hidden before the recombination time, where the
L--T model does not apply. The rest of the paper is devoted to investigating
blueshifts in one specific L--T model, called L--T$(t_B)$ -- the one that
duplicates the luminosity distance vs. redshift relation of the $\Lambda$CDM
model using nonconstant $t_B(r)$ alone. The location of the blueshift-generating
region in the L--T$(t_B)$ spacetime is determined. Profiles of
redshift/blueshift along several rays intersecting the past light cone of the
present central observer are calculated. The L--T$(t_B)$ model matched to
Friedmann is considered, and profiles of redshift/blueshift in such a composite
model are calculated. The requirement of invisibility of blueshifts makes the
L--T$(t_B)$ model astrophysically unacceptable if it should apply back to the
recombination time, but does not ``rule out'' a general L--T model -- it only
puts limits on $\dril {t_B} r$.
\end{abstract}

\maketitle

\section{Motivation and background}\label{intro}

\setcounter{equation}{0}

It was argued \cite{Zibi2011} that the Lema\^{\i}tre \cite{Lema1933} -- Tolman
\cite{Tolm1934} (L--T) models with nonconstant bang-time function $t_B(r)$ are
``ruled out'' because of spectral distortions of light received by the present
central observer that such $t_B$ would cause. The spectral distortions are
expected to arise when blueshifted rays emitted close to those points of the Big
Bang (BB) where $\dril {t_B} r \neq 0$ intersect the past light cone of the
present central observer (PCPO). Investigating the blueshifts and possible
spectral distortions caused by them is a valid research problem. However,
although Ref.\ \cite{Zibi2011} arose out of criticism of the so-called ``void
models of acceleration'', the particular L--T model considered there did not
correspond to any real situation in the Universe (see Appendix
\ref{critique}).\footnote{The author of Ref.\ \cite{Zibi2011} claimed that he
ruled out L--T models ``with significant decaying mode contribution today''. But
the further comments leave the reader with the impression that the whole L-T
class was killed, even though the assumptions actually made ($E = 0$ and a
specific $t_B(r)$ profile) were strongly restricting.}

In the present paper, the blueshifts are investigated in a general L--T model,
and in the model that was derived in Ref.\ \cite{Kras2014} by the method
introduced in Ref.\ \cite{INNa2002}. In the latter, called here the L--T$(t_B)$
model, the accelerated expansion of the Universe is simulated using a suitably
(numerically) constructed nonconstant $t_B(r)$, with the other L--T free
function, $E(r)$, having the same form as in the Friedmann model: $E = -k
r^2/2$, where $k$ is a constant. The equations that ensure the simulation imply
a unique value of $k < 0$, see Sec.\ \ref{LTintro}.

In an L--T model with nonconstant $t_B$, blueshifts are generated only close to
the BB \cite{Szek1980,HeLa1984}. Observers who carry out their observations far
from the BB see light from nearby objects being redshifted. However, light rays
emitted from the BB at those points, where $\dril {t_B} r \neq 0$, behave in a
peculiar way. When a radially directed ray of this class is followed back in
time beginning at a late-epoch observer, the redshift along it at first
increases from zero, but reaches a maximum, then decreases to become negative
(i.e. to turn to blueshift) at some point, to finally become $-1$ at the contact
with the BB. The value $z = -1$ is referred to as {\it infinite blueshift}
\cite{Szek1980,HeLa1984}, see Sec.\ \ref{LTintro}. The locus, where the redshift
along radial rays acquires a maximum, will be termed maximum-redshift
hypersurface (MRH). The locus, where the observed redshift along these rays
turns to blueshift, will be termed zero-redshift hypersurface (ZRH).

To display blueshift to the observer, a ray must build up a sufficiently large
blueshift \textit{before} it intersects ZRH, in order to offset the redshift
accumulated in the later part of the path. Thus, along any ray, the MRH is later
than the ZRH. Along radial rays, the MRH is observer-independent (see Sec.\
\ref{zmax}).

The ZRH is defined only with respect to a given family of observers, and is
different for each family. If ZRH is closer to the BB than the last-scattering
hypersurface, then the blueshifts are hidden in the pre-recombination era, where
the zero-pressure L--T models do not apply. This is why it is important to
locate the ZRH in spacetime, relative to those events, where spectral
distortions would be observable; namely, events along the ray emitted from the
last-scattering hypersurface that reaches the central observer at present.

An L--T metric is never meant to be a global model of the whole Universe. Any
given L--T metric is always meant to model a limited part of our spacetime. It
should be matched to a background metric modelling the rest of the Universe, for
example a Friedmann metric. In the case of the L--T models that simulate
accelerated expansion, it makes sense to apply them only out to such distances,
at which the type Ia supernovae (SNIa) are observed. At present, the largest
redshift observed for a type Ia supernova is $z \approx 1.9$ \cite{Jone2014};
for the supernovae included in the two original projects
\cite{Ries1998,Perl1999}, the largest $z$ was 0.83 \cite{Perl1999}.

The plan of the paper is as follows. Section \ref{LTintro} recalls basic
properties of the L--T models. In Sec.\ \ref{genLT}, it is shown that in a
general L--T model, $\left|\dril {t_B} r\right|$ can always be made sufficiently
small to hide the blueshifts in the pre-recombination epoch, where the
assumption of zero pressure is not realistic, so the L--T model cannot apply.

The rest of the paper is devoted to the L--T($t_B$) model only. In Sec.\
\ref{zmax}, the MRH of the central observer is determined and displayed. In
Sec.\ \ref{numcone}, the profiles of redshift are calculated along a few
characteristic radial rays intersecting  the PCPO. In Sec.\ \ref{matching}, the
L--T($t_B$) model is matched to the Friedmann model across the matter world-tube
that intersects the PCPO at $z = 0.83$, and profiles of redshift/blueshift along
a few characteristic light cones are displayed. Section \ref{conclu} contains
conclusions and a summary. In Appendix \ref{critique}, deficiencies of the model
used in Ref.\ \cite{Zibi2011} are presented.

\section{The Lema\^{\i}tre -- Tolman models}\label{LTintro}

\setcounter{equation}{0}

This is a summary of basic facts about the L--T models. For extended expositions
see Refs.\ \cite{Kras1997,PlKr2006}.

\subsection{General facts}

The numbering of the coordinates will be $(x^0,$ $x^1,$ $x^2,$ $x^3) = (t, r,
\vartheta, \varphi)$ and the signature will be $(+ - - -)$. The metric of the
model is
\begin{equation}\label{2.1}
{\rm d} s^2 = {\rm d} t^2 - \frac {{R_{,r}}^2}{1 + 2E(r)}{\rm d} r^2 -
R^2(t,r)({\rm d}\vartheta^2 + \sin^2\vartheta \, {\rm d}\varphi^2),
\end{equation}
where $E(r)$ is an arbitrary function. The source in the Einstein equations is
dust, i.e. a pressureless fluid. The (geodesic) velocity field of the dust is
\begin{equation}\label{2.2}
u^{\alpha} = {\delta^{\alpha}}_0.
\end{equation}
The function $R(t, r)$ is determined by
\begin{equation}\label{2.3}
{R_{,t}}^2 = 2E(r) + 2M(r) / R,
\end{equation}
$M(r)$ being another arbitrary function; we neglect the cosmological constant.
Throughout this paper only expanding models ($R,_t > 0$) will be considered. The
solutions of (\ref{2.3}) may be written as follows:

When $E > 0$:
\begin{eqnarray}\label{2.4}
R(t,r) &=& \frac M {2E} (\cosh \eta - 1), \nonumber \\
\sinh \eta - \eta &=& \frac {(2E)^{3/2}} M \left[t - t_B(r)\right];
\end{eqnarray}

When $E = 0$:
\begin{equation}\label{2.5}
R(t,r) = \left\{\frac 9 2 M(r) \left[t - t_B(r)\right]^2\right\}^{1/3};
\end{equation}

When $E(r) < 0$:
\begin{eqnarray}\label{2.6}
R(t,r) &=& - \frac M {2E} (1 - \cos \eta), \nonumber \\
\eta - \sin \eta &=& \frac {(-2E)^{3/2}} M \left[t - t_B(r)\right];
\end{eqnarray}
where $t_B(r)$ is one more arbitrary function called the bang time. The Big Bang
occurs at $t = t_B(r)$.

The mass density is
\begin{equation}  \label{2.7}
\kappa \rho = \frac {2{M_{,r}}}{R^2R_{,r}}, \qquad \kappa \df \frac {8\pi G}
{c^2}.
\end{equation}

Equations (\ref{2.1}) -- (\ref{2.7}) are covariant with the transformations $r
\to r' = f(r)$, which may be used to give one of the functions $(M, E, t_B)$ a
handpicked form, in the range where it is monotonic. In this paper, $M,_r > 0$
is assumed, and the following choice of $r$ is made:
\begin{equation}\label{2.8}
M = M_0 r^3,
\end{equation}
where $M_0 > 0$ is an arbitrary constant. The transformations $r = Cr'$, with $C
=$ constant, are still allowed, and they redefine $M_0$ by $M_0 = M'_0 / C^3$.
So, we can assume $M_0 = 1$. Note that $M_0$ has the dimension of length and
represents mass, so the choice of its value amounts to choosing a unit of mass
-- see Sec.\ \ref{numerunits}.

A radial null geodesic is determined by the equation
\begin{equation}\label{2.9}
\dr t r = \pm \frac {R_{,r}} {\sqrt{1 + 2E(r)}},
\end{equation}
where ``$+$'' applies to future outward-directed and past inward-directed
geodesics, and ``$-$'' to the remaining ones. The solution of (\ref{2.9}) is
denoted $t = t_{\rm ng}(r)$. The redshift $z(r)$ along $t_{\rm ng}(r)$ is given
by \cite{Bond1947}, \cite{PlKr2006}
\begin{equation}\label{2.10}
\frac 1 {1 + z}\ \dr z r = \left[ \frac {R_{,tr}} {\sqrt{1 + 2E}} \right]_{\rm ng}.
\end{equation}

At the contact with the BB, null geodesics displayed in the comoving coordinates
have horizontal tangents at those points, where $\dril {t_B} r = 0$, and have
vertical tangents elsewhere. In the first case, $z \to \infty$ at the BB; in the
second case, $z = -1$ at the BB, which is referred to as \textit{infinite
blueshift} \cite{Szek1980,HeLa1984}. Indeed, $z < 0$ means that (frequency
observed) $>$ (frequency at emission), so (frequency observed) $\to \infty$ when
$z \to -1$ and the frequency emitted is finite. However, it should be noted that
a vertical tangent to a light ray at the BB, in comoving coordinates, means that
matter particles are ejected from the BB with the velocity of light, i.e., a
comoving observer at the BB would see zero frequency of the emitted light. So,
some interpretation work is required to decide what an infinite blueshift
actually means: magnifying a finite frequency to an infinitely hard blow at the
observer, or shifting an unobservably soft radiation to the visible range. In
all the Friedmann models (which are subcases of L--T), since $t_B$ is constant,
$z$ is infinite at the BB.

In a general L--T model we have (\cite{PlKr2006}, eqs. (18.104) and (18.112)):
\begin{eqnarray}\label{2.11}
R,_r &=& \left(\frac {M,_r} M - \frac {E,_r} E\right)R \\
&+& \left[\left(\frac 3 2 \frac {E,_r} E - \frac {M,_r} M\right) \left(t -
t_B\right) - t_{B,r}\right] R,_t \nonumber
\end{eqnarray}
when $E \neq 0$, and
\begin{equation}\label{2.12}
R,_r = \frac {M,_r} {3M}\ R - \sqrt{\frac {2M} R} t_{B,r}
\end{equation}
when $E = 0$.

Given a past-directed $t_{\rm ng}(r)$ and $z(r)$, the luminosity distance
$D_L(z)$ of a light source from the central observer is \cite{Cele2000,BKHC2010}
\begin{equation}\label{2.13}
D_L(z) = (1 + z)^2 R\left(t_{\rm ng}(r), r\right).
\end{equation}
The model of Refs.\ \cite{INNa2002} and \cite{Kras2014}, further investigated
here, was constructed so that $D_L(z)$, calculated along the PCPO, is the same
as in the $\Lambda$CDM model:
\begin{equation}\label{2.14}
D_L(z) = \frac {1 + z} {H_0} \int_0^z \frac {{\rm d} z'} {\sqrt{\Omega_m (1 +
z')^3 + \Omega_{\Lambda}}},
\end{equation}
where $H_0$ is related to the Hubble ``constant'' ${\cal H}_0$ by $H_0 = {\cal
H}_0 / c$, and $(\Omega_m, \Omega_{\Lambda}) = (0.32, 0.68)$ are parameters
defined by observations \cite{Plan2013}; see Sec.\ \ref{numerunits}.

Note that the duplication of $D_L(z)$ occurs only along a single light cone.
Observations that are sensitive to the dynamics of the model, for example
redshift drift \cite{QABC2012}, could distinguish between the $\Lambda$CDM and
L--T models having the same $D_L(z)$ at present.

The $R_{\rm ng}(r) \df R\left(t_{\rm ng}(r), r\right)$ in (\ref{2.13}) is the
angular diameter distance, and it is not an increasing function of $r$. At the
intersection with the hypersurface $t(r)$, implicitly determined by the equation
\cite{KrHe2004b}
\begin{equation}\label{2.15}
R = 2M,
\end{equation}
the function $R_{\rm ng}(r)$ acquires a maximum, and becomes decreasing for
greater $r$. The hypersurface determined by (\ref{2.15}) is called
\textit{apparent horizon} (AH). It is a difficult obstacle to numerical
calculations because several quantities become 0/0 there, see Refs.\
\cite{Kras2014,Kras2014b} and further references cited in them. Traces of those
difficulties will appear here in a few graphs as numerical instabilities. The
values of $r$ and $z$ at the intersection of the PCPO with the AH in the
L--T($t_B$) model are \cite{Kras2014}
\begin{equation}\label{2.16}
\left(\begin{array}{l}
 r \\
 z \\
\end{array}\right)_{\rm AH} =
\left(\begin{array}{l}
0.3105427968086945 \\
1.582430687623614 \\
\end{array}\right).
\end{equation}

For technical reasons, the $t(r)$ and $z(r)$ curves crossing the point $r =
r_{\rm AH}$ were calculated separately in the ranges $r < r_{\rm AH}$ and $r >
r_{\rm AH}$, as explained in Refs.\ \cite{Kras2014} and \cite{Kras2014b}.
Therefore, they are differently coloured in each of these ranges.

\subsection{The L--T model with $2E = - k r^2$ that obeys
(\ref{2.14})}\label{LTwithnonzeroE}

This is the L--T($t_B$) model. In it we have \cite{Kras2014}
\begin{equation}\label{2.17}
2E = - kr^2,
\end{equation}
where $k < 0$ is a constant. This $E$ is the same as in the $k < 0$ Friedmann
model. Numerical fitting of the solution of (\ref{2.10}) to the values of $(r,
z)$ at $(0, 0)$ and at the AH determined the value of $k$ \cite{Kras2014},
\begin{equation}\label{2.18}
k = - 4.7410812.
\end{equation}
{}From (\ref{2.9}) and (\ref{2.17}) we have on a light cone
\begin{equation}\label{2.19}
\dr t r = \pm \frac {R,_r} {\sqrt{1 - k r^2}}.
\end{equation}
Using (\ref{2.8}) and (\ref{2.17}) we get from (\ref{2.11})
\begin{equation}\label{2.20}
R,_r = \frac R r - r t_{B,r} \sqrt{\frac {2M_0 r} R - k}.
\end{equation}
With (\ref{2.17}), eqs. (\ref{2.4}) become
\begin{eqnarray}
\cosh \eta &=& 1 - \frac {k R} {M_0 r}, \label{2.21} \\
t - t_B &=& \frac {M_0} {(- k)^{3/2}} (\sinh \eta - \eta). \label{2.22}
\end{eqnarray}
Using (\ref{2.17}) and (\ref{2.20}) -- (\ref{2.22}), the set of equations
\{(\ref{2.19}), (\ref{2.22}), (\ref{2.10})\} was numerically solved in Ref.\
\cite{Kras2014} for $t(r)$, $t_B(r)$ and $z(r)$ along the PCPO. The tables
representing those solutions will be used here.

The L--T($t_B$) model is determined around the center of symmetry up to those
worldlines of dust that leave the BB at its contact with the PCPO. Extensions
beyond the world-tube composed of those worldlines are possible, but are not
determined by (\ref{2.13}) -- (\ref{2.14}) and are not considered here. This
model need not be used in this full range. A subset can be cut out of it and
matched to a background model along a narrower world-tube.

\subsection{The numerical units}\label{numerunits}

The following values are assumed here:
\begin{equation}\label{2.23}
(\Omega_m, \Omega_{\Lambda}, H_0, M_0) = (0.32, 0.68, 6.71, 1)
\end{equation}
the first two after Ref.\ \cite{Plan2013}. The $H_0$ is related to the Hubble
constant ${\cal H}_0$ \cite{Plan2013} by

\begin{equation}\label{2.24}
{\cal H}_0 = c H_0 = 67.1\ {\rm km/(s} \times {\rm Mpc}),
\end{equation}
so $H_0$ is measured in 1/Mpc. Consequently, choosing a value for $H_0$ amounts
to defining a numerical length unit (NLU). Our time coordinate is $t = c \tau$,
where $\tau$ is measured in time units, so $t$ is measured in length units. So
it is natural to take the NLU also as the numerical time unit (NTU). Taking for
the conversion factors \cite{unitconver}
\begin{eqnarray}\label{2.25}
1\ {\rm pc} &=& 3.086 \times 10^{13}\ {\rm km}, \nonumber \\
1\ {\rm y} &=& 3.156 \times 10^7\ {\rm s},
\end{eqnarray}
the following relations result:
\begin{eqnarray}\label{2.26}
&& 1\ {\rm NTU} = 1\ {\rm NLU} = 3 \times 10^4\ {\rm Mpc} \nonumber \\
&&= 9.26 \times 10^{23}\ {\rm km} = 9.8 \times 10^{10}\ {\rm y}.
\end{eqnarray}
The age of the Universe inferred from observations is \cite{Plan2013}
\begin{equation}\label{2.27}
T = 13.819 \times 10^9\ {\rm y} = 0.141\ {\rm NTU}.
\end{equation}

As already mentioned below (\ref{2.8}), $M_0$ represents mass, but has the
dimension of length ($M_0 = G m_0/c^2$, where $m_0$ is measured in mass units).
The choice $M_0 = 1$ NLU made in (\ref{2.23}) implies the mass unit $M_0 c^2/G
\approx 10^{54}$ kg, but it will not appear in any other way than via $M_0$.

\section{The maximum-redshift hypersurface in a general L--T model}\label{genLT}

\setcounter{equation}{0}

Consider a radial light ray $t_{\rm ng}(r)$ reaching a comoving observer at a
sufficiently late time (below, it will become clear what ``sufficiently late''
means). When we follow that ray back in time and calculate the redshift $z(r)$
along it, then, initially, $z$ increases from $z = 0$ at the observation event.
However, if the ray was emitted at the BB at such a point, where $\dril {t_B} r
\neq 0$, then $z(r)$ will reach a maximum somewhere in the past, and will then
decrease, to become $z = -1$ at the intersection of the ray with the BB. The
maximum redshift is achieved where $\dril z r = 0$, i.e., from (\ref{2.10}),
where $R,_{tr} = 0$. The hypersurface determined by $R,_{tr} = 0$ is
observer-independent; this is the MRH described in Sec.\ \ref{intro}. From
(\ref{2.11}), using (\ref{2.3}), we have
\begin{eqnarray}\label{3.1}
R,_{tr} &=& \frac {E,_r} {2E}\ R,_t - \frac M {R^2}\ \left(\frac 3 2 \frac
{E,_r} E - \frac {M,_r} M\right) \left(t - t_B\right) \nonumber \\
&+& \frac M {R^2}\ t_{B,r}
\end{eqnarray}
when $E\neq 0$, and
\begin{equation}\label{3.2}
R,_{tr} = \left(\frac {M,_r} {3M} + \frac {\sqrt{2M}} {R^{3/2}}\ t_{B,r}\right)
R,_t
\end{equation}
when $E = 0$. From now on, the cases $E > 0$ and $E < 0$ have to be considered
separately.

\medskip

{\underline {\bf $E > 0$}}

\medskip

Using (\ref{2.3}), (\ref{2.4}) and (\ref{3.1}), the equation $R,_{tr} = 0$ is
rewritten as
\begin{equation}\label{3.3}
\left(t - t_B\right) \left[\frac {E,_r} {2E}\ F_1(\eta) + \frac 3 r\right] = -
t_{B,r},
\end{equation}
where
\begin{equation}\label{3.4}
F_1(\eta) \df \frac {\sinh \eta (\cosh \eta - 1)} {\sinh \eta - \eta} - 3.
\end{equation}
The conditions for no shell crossings in the case $E > 0$ are \cite{HeLa1985},
\cite{PlKr2006}
\begin{equation}\label{3.5}
E,_r > 0, \qquad t_{B,r} < 0.
\end{equation}
Hence, in a region with no shell crossings,\footnote{It has to be recalled that
in the L--T model that reproduces (\ref{2.14}) with $t_{B,r} \equiv 0$, a region
with shell crossings does exist \cite{Kras2014b,Kras2014c}. This is why we
cannot assume that the whole model is free of shell crossings; it is to be
expected that shell crossings will also exist when $\left|t_{B,r}\right|$ is
small but nonzero.} the right-hand side of (\ref{3.3}) is positive, and so is
the coefficient of $F_1(\eta)$. We also have
\begin{equation}\label{3.6}
F_1(\eta) > 0, \qquad \dril {F_1} {\eta} > 0
\end{equation}
for all $\eta > 0$. From this follows

\bigskip

{\bf Lemma 3.1}

\medskip

\noindent For every $\varepsilon > 0$ there exists a $\delta > 0$ such that $t -
t_B < \varepsilon$ if $\left|t_{B,r}\right| < \delta$.

\bigskip

\noindent The proof is given in Appendix \ref{provelem3.1}.

Consequently, by choosing $\left|t_{B,r}\right|$ sufficiently small, $t$ can be
made arbitrarily close to $t_B$; in particular, earlier than the recombination
time. Thus, the MRH can be hidden in the pre-recombination epoch, where the
zero-pressure L--T models cannot apply, and blueshifts will not arise in the
L--T region.

Note that (\ref{3.3}) -- (\ref{3.6}) imply that the MRH does not exist if
$t_{B,r} = 0$ everywhere. (Formally, (\ref{3.3}) implies then $t = t_B$, but we
know from elsewhere \cite{Szek1980,HeLa1984} that in this case $z \to \infty$ at
the BB, so $\dril z r \to \infty$, too.) This holds, for example, in the L--T
model of Ref.\ \cite{Kras2014b}.

\newpage

{\underline {\bf $E = 0$}}

\medskip

Then, using (\ref{3.2}), the equation $R,_{tr} = 0$ implies
\begin{equation}\label{3.7}
R = 2^{1/3} M_0 r^{5/3} \left(- t_{B,r}\right)^{2/3}.
\end{equation}
The no-shell-crossing condition here is $t_{B,r} < 0$. Thus, by choosing
$\left|t_{B,r}\right|$ sufficiently small, we can make $R$ arbitrarily close to
zero, i.e. to the BB. So, again, the blueshifts can be removed from the L--T
region.\footnote{But the model with $E = 0$ cannot obey (\ref{2.14})
\cite{Kras2014}.}

\medskip

{\underline {\bf $E < 0$}}

\medskip

Using (\ref{2.6}) and (\ref{3.1}) and assuming $0 \leq \eta \leq \pi$ (the
Universe is expanding), $R,_{tr} = 0$ is rewritten as
\begin{equation}\label{3.8}
\left(t - t_B\right) \left[\frac {E,_r} {2E}\ G_1(\eta) + \frac 3 r\right] = -
t_{B,r},
\end{equation}
where
\begin{equation}\label{3.9}
G_1(\eta) \df \frac {\sin \eta (1 - \cos \eta)} {\eta - \sin \eta} - 3.
\end{equation}
In this case, the analogue of Lemma 3.1 cannot be proved, since the function
$G_1$ is negative and decreasing for all $0 < \eta < \pi$, while the
no-shell-crossing conditions do not imply a unique sign for
$E,_r/E$.\footnote{Since $E(0) = 0$ must hold in order to avoid a permanent
central singularity \cite{PlKr2006}, and $E(r) < 0$ at $r > 0$ by assumption in
this case, the consequence is $E,_r < 0$ in a vicinity of $r = 0$, i.e., $E,_r/E
> 0$ for small $r$. However, this argument does not hold for large $r$. The
no-shell-crossing  conditions require only that $t_{B,r} < 0$ and $2 \pi
\left(\frac 3 2 \frac {E,_r} E - \frac {M,_r} M\right) - \frac {(- 2E)^{3/2}} M
t_{B,r} < 0$ \cite{PlKr2006}.} So, this case was considered for completeness
only. It is known \cite{Kras2014b} that with $t_{B,r} \equiv 0$ the relation
(\ref{2.13}) -- (\ref{2.14}) can be duplicated in an L--T model only with $E >
0$ at all $r > 0$. Thus, it is to be expected that the model with $E < 0$ at $r
> 0$ will be inapplicable also when $\left|t_{B,r}\right|$ is small but nonzero.

\section{The maximum-redshift hypersurface in the L--T($t_B$) model}\label{zmax}

\setcounter{equation}{0}

In the L--T($t_B$) model, the nonconstant $t_B(r)$ is uniquely (numerically)
determined by (\ref{2.10}), so $\dril {t_B} r$ is also fixed. Consequently, the
method of removing blueshifts from the L--T epoch, presented in Sec.\
\ref{genLT}, cannot be applied here. As will be seen, blueshifts in  this model
occur later than the recombination epoch in a large region around the center. A
radical solution of the problems with blueshifts would be to assume that the
L--T($t_B$) model applies only as far back in time as $\dril z r > 0$ along
radial rays. However, it is useful to know exactly where the
blueshift-generating region is located and how the blueshifts would make
themselves visible to a late-time observer in this model. These questions will
be dealt with in the remaining part of this paper.

As in Sec.\ \ref{genLT}, the location of the MRH in spacetime is determined by
the equation $R,_{tr} = 0$. However, caution is required in interpreting the
solution. Equation (\ref{2.10}) shows that $R,_{tr}$ might vanish also at those
points, where $z \to \infty$. See below for more on this.

Using (\ref{2.20}) for $R,_r$, the equation $R,_{tr} = 0$ is
\begin{equation}\label{4.1}
\sqrt{\frac {2M_0 r} R - k} = - \frac {M_0 r^3 t_{B,r}} {R^2},
\end{equation}
where (\ref{2.3}), (\ref{2.17}) and (\ref{2.8}) were used to eliminate $R,_t$.
With $t_{B,r} < 0$ and $k < 0$, (\ref{4.1}) is solvable and implicitly defines
(via $R(t,r)$) the $t(r)$ function along the MRH.

For numerical handling, it is more convenient to square (\ref{4.1}) and
substitute for $R$ from (\ref{2.21}), obtaining
\begin{equation}\label{4.2}
x^4 + x^3 + k^3 \left(\frac {r t_{B,r}} {4 M_0}\right)^2 = 0,
\end{equation}
where
\begin{equation}\label{4.3}
x \df \sinh^2 (\eta/2).
\end{equation}
Where $r > 0$ and $t_{B,r} < 0$, the solution of (\ref{4.2}) obeys
\begin{equation}\label{4.4}
0 < x < x_{\rm max} \df - k \ \left(\frac {r t_{B,r}} {4 M_0}\right)^{2/3}.
\end{equation}
However, (\ref{4.2}) implies $x = 0$ at those $r$, where $t_{B,r} = 0$. This
means $\eta = 0$, i.e. $R = 0$. This is the BB, where $z \to \infty$. The
conclusion is that the MRH does not exist along those rays that hit the BB where
$t_{B,r} = 0$. Also, (\ref{4.2}) implies $x = 0$ at $r = 0$. The point
determined by $x = 0$ ($\Longrightarrow \eta = 0$) and $r = 0$ is the central
point of the BB, where $R = 0$. But (\ref{4.2}) was obtained from (\ref{4.1}) by
squaring and multiplying by $R^4$. Consequently, the solution of (\ref{4.1}) is
not determined at $r = 0$, although the limit $r \to 0$ of the solution found at
$r
> 0$ may exist.

Having found (numerically) $x(r)$, and thus also $\eta(r)$ from (\ref{4.3}), we
find $t(r)$ on the MRH from (\ref{2.4}):
\begin{equation}\label{4.5}
t_{\rm MRH}(r) = t_B(r) + \frac {M_0} {(-k)^{3/2}}\ \{\sinh [\eta(r)] -
\eta(r)\}.
\end{equation}

Equation (\ref{4.2}) was derived assuming that the null geodesics, on which
$z(r)$ is calculated, are radial. But it makes no reference to the initial point
of the geodesic arc. Consequently, the MRH is observer-independent. The maximal
\textit{value} of redshift will depend on the initial point, where $z = 0$, but
the \textit{location} of the maximum will not: the maximum along a given
geodesic will occur always at the same $r$.

In order to use (\ref{4.5}), we need to know the function $t_B(r)$. It was
numerically calculated in Ref.\ \cite{Kras2014}, but only up to $r \approx
1.05584$ (corresponding to $t_B \approx -0.139454$ NTU), which is not sufficient
for the present purpose. Consequently, it had to be re-calculated and extended,
and the way of extending needs an explanation.

\begin{figure}[h]
\hspace{-1cm}
\includegraphics[scale = 0.5]{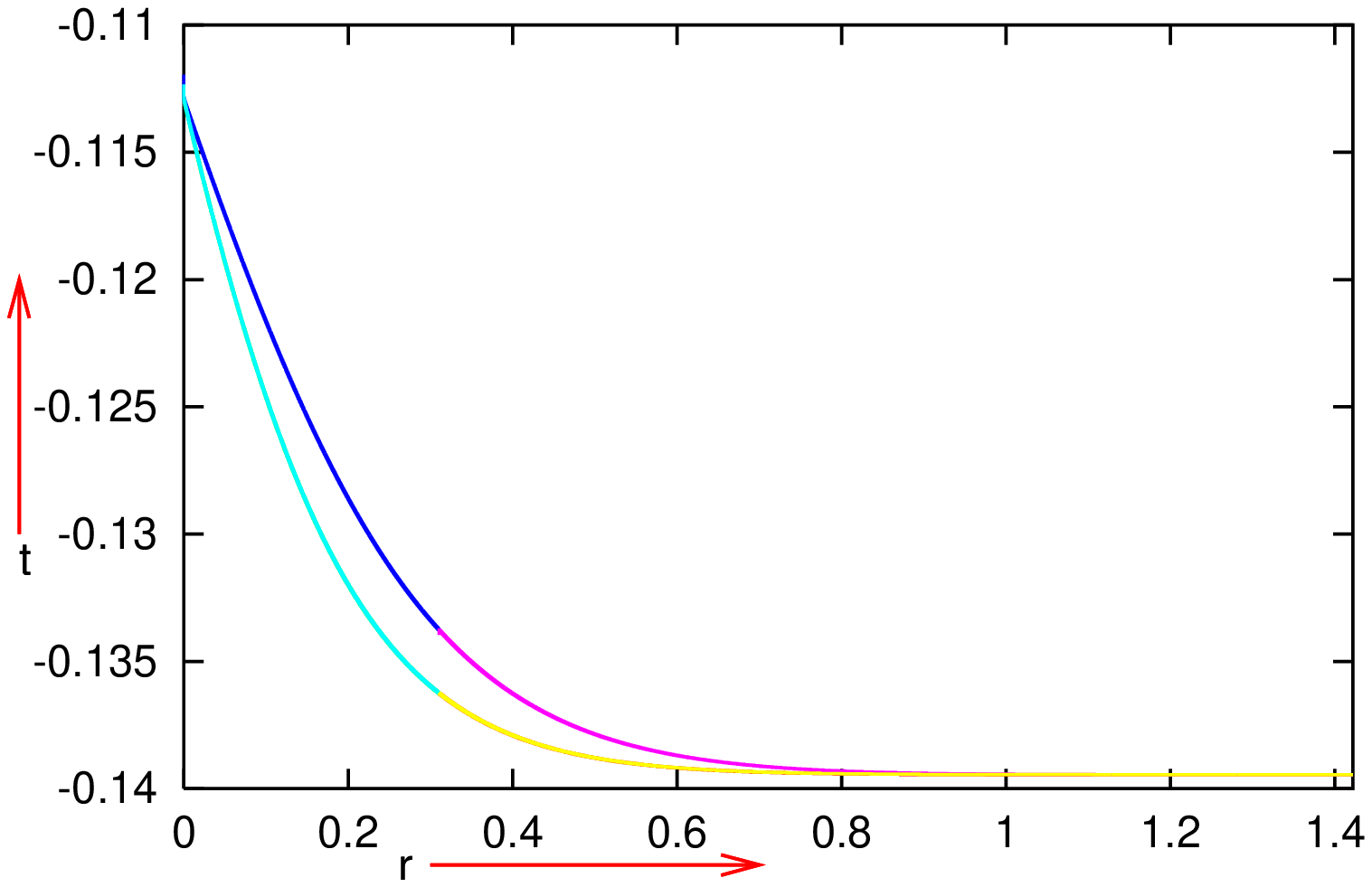}
${ }$ \\[-4.7cm]
\hspace{1.4cm}
\includegraphics[scale = 0.32]{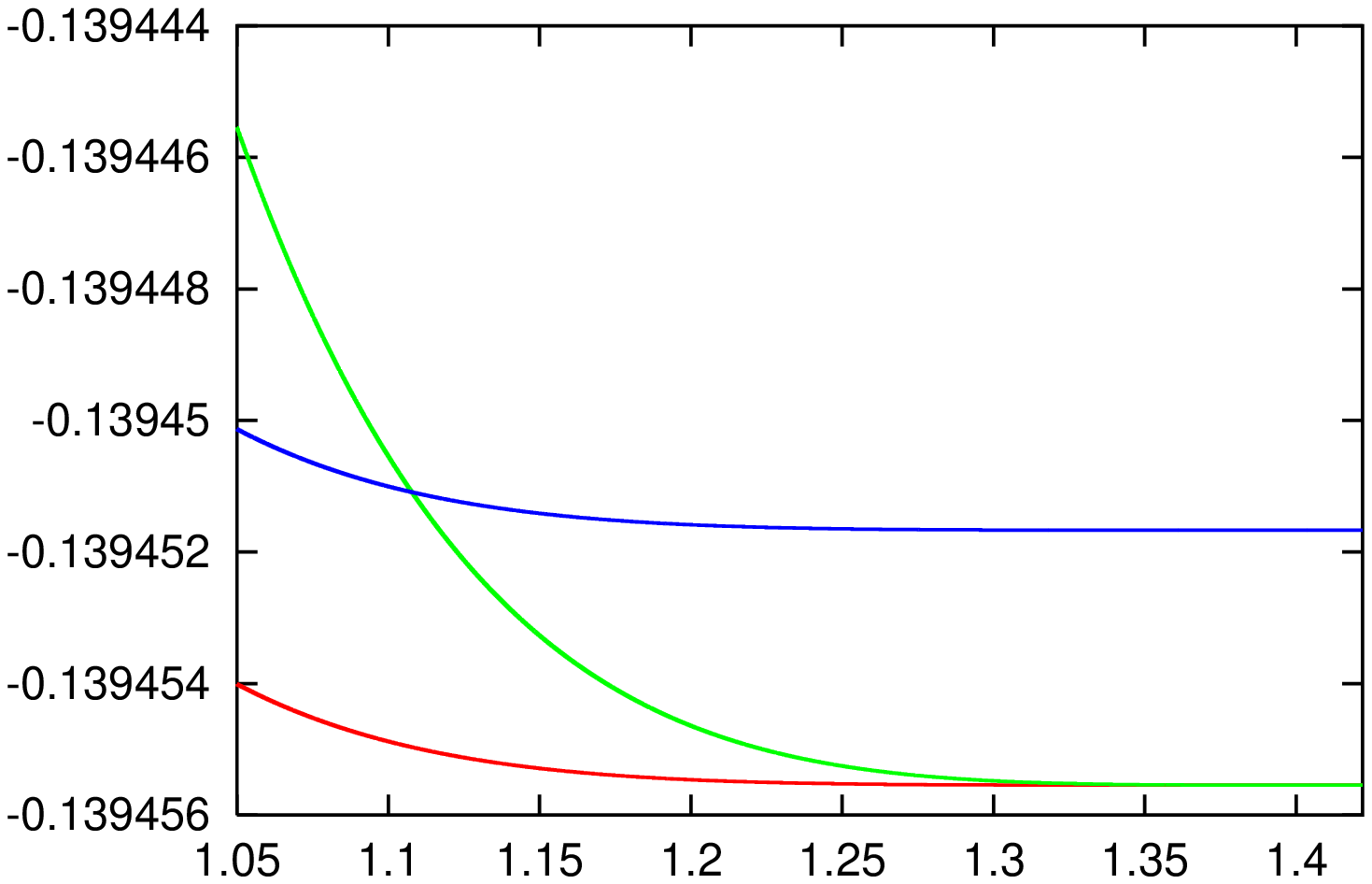}
\vspace{1.5cm}
\caption{{\bf Main panel:} The functions $t_B(r)$ (lower curve) and $t_{\rm
MRH}(r)$. The $t_B(r)$ acquires the constant value $t_{Bc}$ given by (\ref{4.7})
on approaching the right end. {\bf Inset:} Closeup view of the region, where
$t_B(r)$ (the lowest curve) and $t_{\rm MRH}(r)$ (the curve with the uppermost
left end) become tangent. The third curve is the $t_{\rm rec}(r)$ of
(\ref{4.8}). See text for more explanation.} \label{drawtbandzz}
\end{figure}

The numerical step was $\Delta r \approx 5 \times 10^{-6}$. Beginning at
\begin{equation}\label{4.6}
r \df r_c = 1.4131983072777050,
\end{equation}
the numerically calculated $t_B$ became constant,
\begin{equation}\label{4.7}
t \df t_{Bc} = -0.13945554689046649\ {\rm NTU},
\end{equation}
and this value was maintained from step $n = 221923$ for the next 1757 steps
(the calculation broke down at $r \df r_f = 1.4219332552803152$, with the
Fortran program saying that $t_B =$ NaN for all $r > r_f$). So, it was assumed
that $t_B(r) = t_{Bc}$ at the contact with the PCPO, and the $t_B(r)$ curve was
extended ``by hand'' as the straight line $t = t_{Bc}$. The extended graph of
$t_B(r)$ is shown in Fig. \ref{drawtbandzz} together with $t_{\rm MRH}(r)$.
Given the table of values of $t_B(r)$, the $t_{B,r}(r)$ needed to solve
(\ref{4.2}) is easy to calculate.

The inset in Fig. \ref{drawtbandzz} includes the graph of the recombination
time, given approximately by \cite{swinxxxx}
\begin{equation}\label{4.8}
t_{\rm rec}(r) - t_B(r) = 3.8 \times 10^5\ {\rm y} = 3.88 \times 10^{-6}\ {\rm
NTU}.
\end{equation}
This will be used for illustration only. The correct $t_{\rm rec}(r)$ would have
to be calculated by determining the $t - t_B$, at which the density in our model
becomes equal to the density at recombination in the $\Lambda$CDM model.
However, this more exact calculation would introduce only a small correction to
(\ref{4.8}), which would not substantially improve the usefulness of it. As
further calculations will show, along most rays both the MRH and the ZRH will
occur much later than the time given by (\ref{4.8}). At the scale of the main
panel of Fig. \ref{drawtbandzz}, the graph of $t_{\rm rec}(r)$ is
indistinguishable from the graph of $t_B(r)$.

The following facts about Fig. \ref{drawtbandzz} need to be noted:

1. The right end of the graphs is at $r = 1.422$. This is where $z(r)$ along the
PCPO was found to be unmanageably large ($z \approx 1.6237 \times 10^{229}$)
\cite{Kras2014}, signaling the near-contact of the PCPO with the BB.

2. The $t_{\rm rec}(r)$ and $t_{\rm MRH}(r)$ curves intersect at $r \df r_x$
$\approx 1.107817$. For $r > r_x$, the MRH is earlier than the hypersurface of
last-scattering, and thus becomes astrophysically irrelevant, as the L--T model
is inadequate for describing the epoch $t < t_{\rm rec}(r)$.

3. The redshift corresponding to $r_x$ is $z_x \approx 57.88$. This is much
larger than $z_{\rm far} = 10$ \cite{McMa2005}, the largest observed redshift
apart from CMB.

\section{Light rays intersecting the past light cone of the present central
observer}\label{numcone}

\setcounter{equation}{0}

The profile of the PCPO calculated in Ref.\ \cite{Kras2014} is shown in the main
panel of Fig. \ref{drawsecconeb}. It becomes tangent to the $t_B(r)$ at $r
\approx 1.42182$. We will now determine the intersections with the ZRH for rays
received by observers sitting on the PCPO at a few characteristic positions.

\subsection{Ray B}\label{rayB}

\begin{figure}[h]
\hspace{-0.5cm}
\includegraphics[scale=0.63]{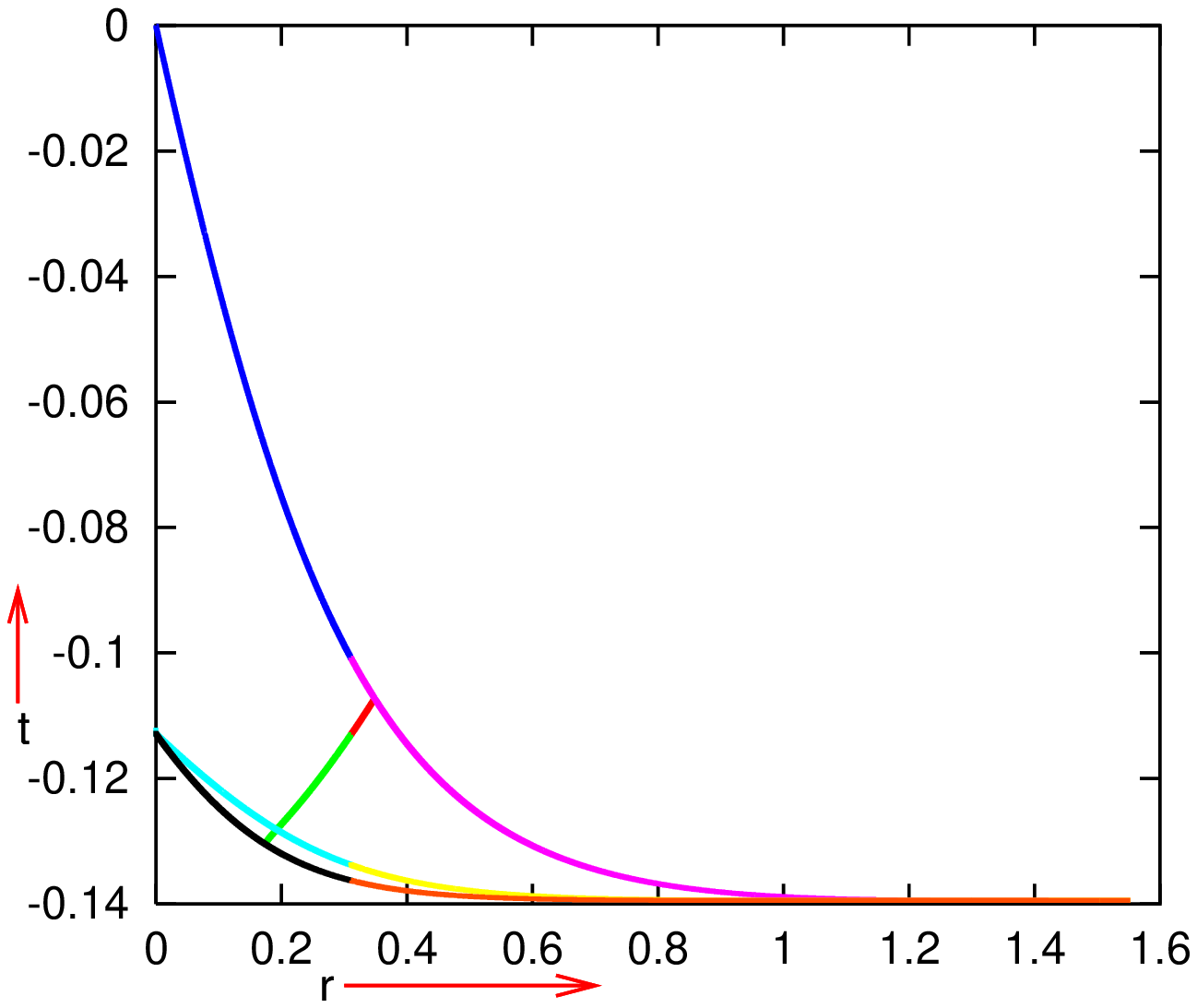}
${ }$ \\[-6.5cm]
\hspace{1.8cm}
\includegraphics[scale=0.5]{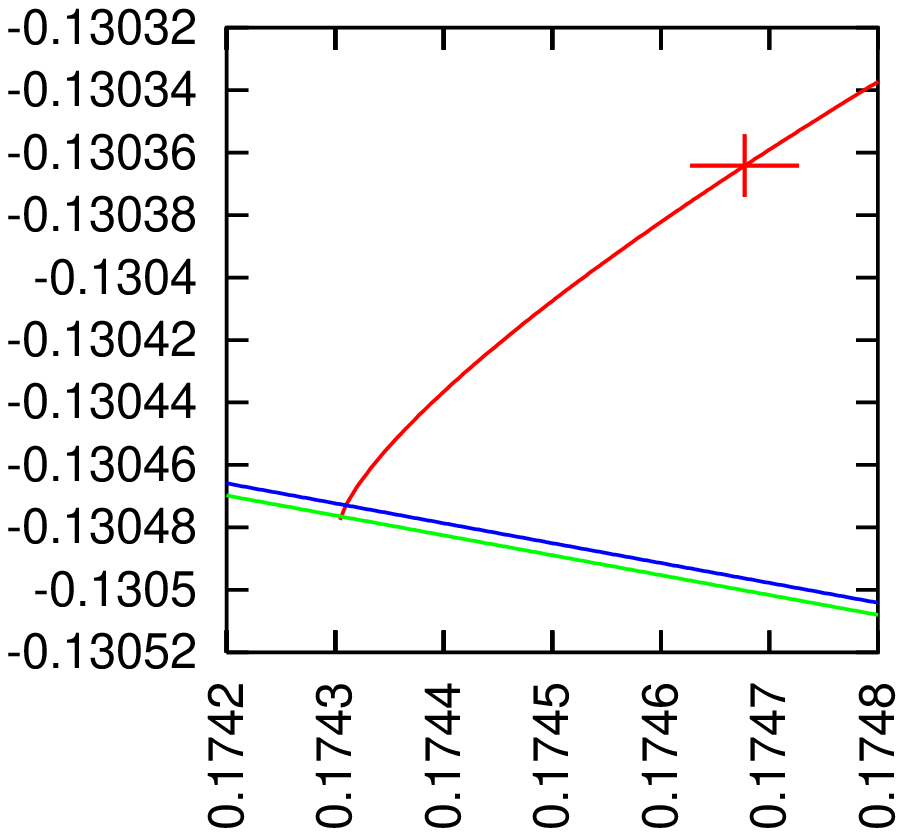}
\vspace{2cm} \caption{{\bf Main panel:} The uppermost curve is the profile of
the past light cone of the present central observer. The other two decreasing
curves are those from the main panel of Fig. \ref{drawtbandzz}. The increasing
curve is ray B. See text for more explanation. {\bf Inset:} Magnified view of
the neighbourhood where $z < 0$ along ray B. The two decreasing lines are
$t_B(r)$ (lower) and $t_{\rm rec}(r)$ of (\ref{4.8}). The increasing curve is
ray B. The cross marks the point where $z = 0$. The $t_{\rm MRH}$ profile is
far above the upper margin.} \label{drawsecconeb}
\end{figure}

Consider the observer O$_{\rm b}$ (``b'' for ``border'') who intersects the PCPO
at $z = z_{\rm fSN} = 1.9$. As noted above, this is the largest observed
redshift corresponding to a supernova of type Ia \cite{Jone2014}. The functions
$z(r)$, $t(r)$ and $R(r)$ along the PCPO were calculated in Ref.\
\cite{Kras2014}. In their tables of values, the $z$ nearest to $z_{\rm fSN}$ and
the corresponding $r$, $t$ and $R$ at the PCPO are
\begin{eqnarray}
z_{\rm b} &=& 1.900028454789241, \label{5.1} \\
r_{\rm b} &=& 0.3486128555616366, \label{5.2} \\
t_{\rm b} &=& -0.10726235253032952, \label{5.3} \\
R_{\rm b} &=& 0.0594055585753355889. \label{5.4}
\end{eqnarray}
Equations (\ref{5.2}) and (\ref{5.3}) define the initial conditions for the
outgoing radial light ray that O$_b$ receives at the moment of intersecting the
PCPO. It was calculated backward from this event and will be called ray B. It is
the increasing curve in Fig. \ref{drawsecconeb}.

\begin{figure}[h]
\begin{center}
\includegraphics[scale=0.6]{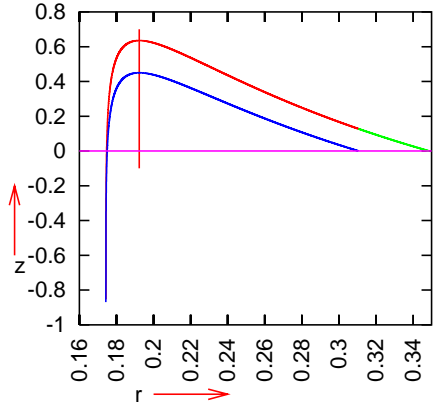}
\caption{Redshift $z(r)$ along ray B, calculated from the initial point at the
past light cone of the present central observer (upper curve) and from the
initial point at the intersection of ray B with $r = r_{\rm AH}$. The vertical
bar marks $r = r_{\rm bmax}$ given by (\ref{5.5}). More explanation in the text.
} \label{drawseczb}
\end{center}
\end{figure}

Figure \ref{drawseczb} shows the graph of $z(r)$ along ray B. The numerical
calculation broke down near the singularity, so the value of $z$ at $t = t_B$
could not be calculated, but it is known to be $-1$ there
\cite{Szek1980,HeLa1984}. The nearest value to $-1$ that was yet calculated was
$z = -0.85255346539197885$, achieved at $r = 0.17430456376783951$. The right end
of the graph is at the $r_{\rm b}$ given by (\ref{5.2}), where the initial value
on ray B was $z = 0$.

The maximal redshift along this ray is $z = 0.63536180471132442$, achieved at
\begin{equation}\label{5.5}
\left(\begin{array}{l}
r \\
t \\
\end{array}\right)_{\rm bmax} = \left(\begin{array}{l}
0.19233778370400151 \\
-0.12814671882472262 \\
\end{array}\right).
\end{equation}
This maximum fits on the $t_{\rm MRH}(r)$ curve from Fig. \ref{drawsecconeb} up
to better than $2 \times 10^{-7}$ NTU = 19 600 y.

For illustration, Fig. \ref{drawseczb} contains also the graph of redshift along
ray B, with the initial value $z = 0$ not at the PCPO, but at the intersection
of ray B with the line $r = r_{\rm AH}$. As predicted, the second maximum is at
a different $z$, but at $r'$, for which $\left|r' - r_{\rm bmax}\right| \approx
3 \times 10^{-6}$.

\subsection{Ray OB}\label{rayOB}

\begin{figure}[h]
\hspace{-0.4cm}
\includegraphics[scale=0.65]{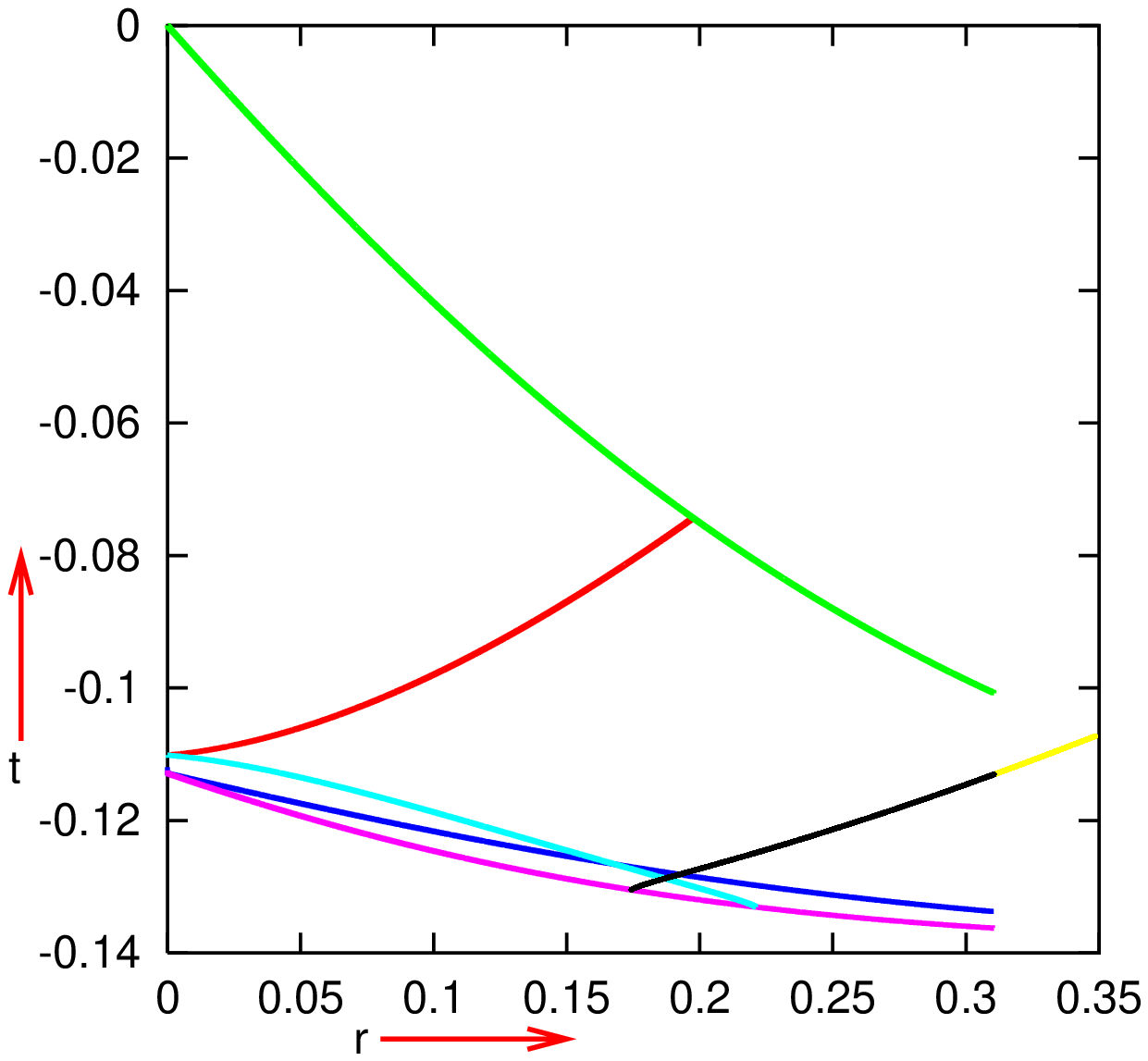}
${ }$ \\[-6.8cm]
\hspace{3.5cm}
\includegraphics[scale=0.35]{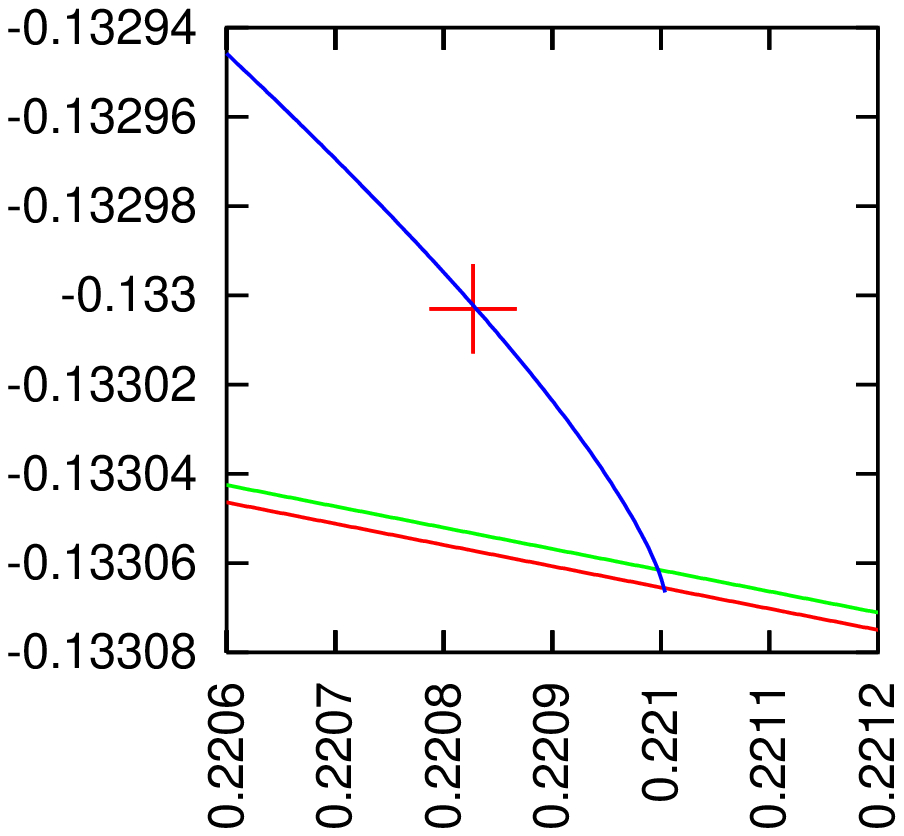}
\vspace{4cm}
\caption{{\bf Main panel:} The uppermost and the two lowest curves are those
from Fig. \ref{drawsecconeb}. The two solid arcs at left represent ray OB. The
dotted arc at right is ray B. See text for more explanation. {\bf Inset:}
Magnified view of the neighbourhood where $z < 0$ along ray OB. The two lowest
lines are the same as in the inset in Fig. \ref{drawsecconeb}. The third curve
is ray OB. The cross marks the point where $z = 0$. The profile of the MRH is
again way above the upper margin. }
\label{drawsecconestaresn}
\end{figure}

Consider the second observer O$_{\rm ob}$ (for ``old border'') intersecting the
PCPO at $z_{\rm ob} = 0.83$, which is the largest SNIa redshift measured in the
twin projects that first reported the accelerated expansion \cite{Perl1999}.
Again, from the $z(r)$, $t(r)$ and $R(r)$ tables in Ref.\ \cite{Kras2014} we
find the $z$ nearest to $z_{ob}$, and the corresponding $r$, $t$ and $R$ on the
PCPO:
\begin{eqnarray}
z_{\rm ob} &=& 0.8300015499642085, \label{5.6} \\
r_{\rm ob} &=& 0.19751142662007609, \label{5.7} \\
t_{\rm ob} &=& -0.0743328307281575784, \label{5.8} \\
R_{\rm ob} &=& 0.0540017311248809709. \label{5.9}
\end{eqnarray}
The ray emitted at the BB and received by O$_{\rm ob}$ at the event given by
(\ref{5.7}) -- (\ref{5.8}) will be denoted OB and is shown in Fig.
\ref{drawsecconestaresn}, together with the other curves from Fig.
\ref{drawsecconeb}. Ray OB goes into the past from the PCPO, reaching the center
$r = 0$ at $t = -0.11014300011521007$ NTU. It goes to the other side of the
center, but its continuation beyond $r = 0$ is drawn in mirror-reflection.

\begin{figure}[h]
\begin{center}
\vspace{5mm}
\includegraphics[scale=0.6]{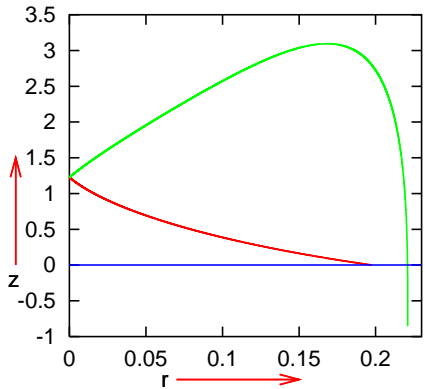}
\caption{Redshift $z(r)$ along ray OB.}
\label{drawzstaresn}
\end{center}
\end{figure}

Figure \ref{drawzstaresn} shows the graph of $z(r)$ along ray OB. The graph
begins at the $r_{\rm ob}$ given by (\ref{5.7}), where $z = 0$, and proceeds to
the left. The curve $z(r)$ hits the center with $z = 1.2266046302084745$ and
goes to the left side of the $z$-axis, but, as before, the figure shows the
mirror-image of the continuation. From this point on, $z(r)$ increases to the
maximum $z = 3.0946480957646290$ attained at
\begin{equation}\label{5.10}
\left(\begin{array}{l}
r \\
t \\
\end{array}\right)_{\rm obmax} = \left(\begin{array}{l}
0.16821171418720421 \\
-0.12662412927673727 \\
\end{array}\right).
\end{equation}
The $t_{\rm obmax}$ given above agrees with the corresponding $t$ on the $t_{\rm
MRH}(r)$ curve to better than $10^{-6}$ NTU $= 9.8 \times 10^4$ y. Beyond this
maximum, $z(r)$ decreases to $z = -0.85628254978650187$ attained at $r =
0.22100398884102759$.

Note that the redshift in Fig. \ref{drawzstaresn} becomes negative at $r \approx
0.2208$, which is larger than the $r_{\rm ob}$, given by (\ref{5.7}). Thus, if
our L--T model were matched to a Friedmann background at $r$ between $r_{\rm
ob}$ and 0.2208, the ray (followed back in time) would enter the Friedmann
region with the redshift still being positive, and would start building up
more-positive redshifts from then on. See more on this in Sec.\ \ref{matching}.

It is interesting that all the qualitative properties of blueshift described
here (blueshift being infinite when $\dril {t_B} r \neq 0$ at the contact of the
ray with the BB, being visible to all observers along the blueshifted ray,
perturbing the CMB spectrum) were mentioned without proof by Szekeres already in
1980 \cite{Szek1980}; he even drew the MRH profile for $E = 0$ and $t_B(r) = 1 /
\left(1 + r^2\right)$.

\subsection{Ray N}\label{rayN}

\begin{figure}[h]
\hspace{-0.7cm}
\includegraphics[scale=0.56]{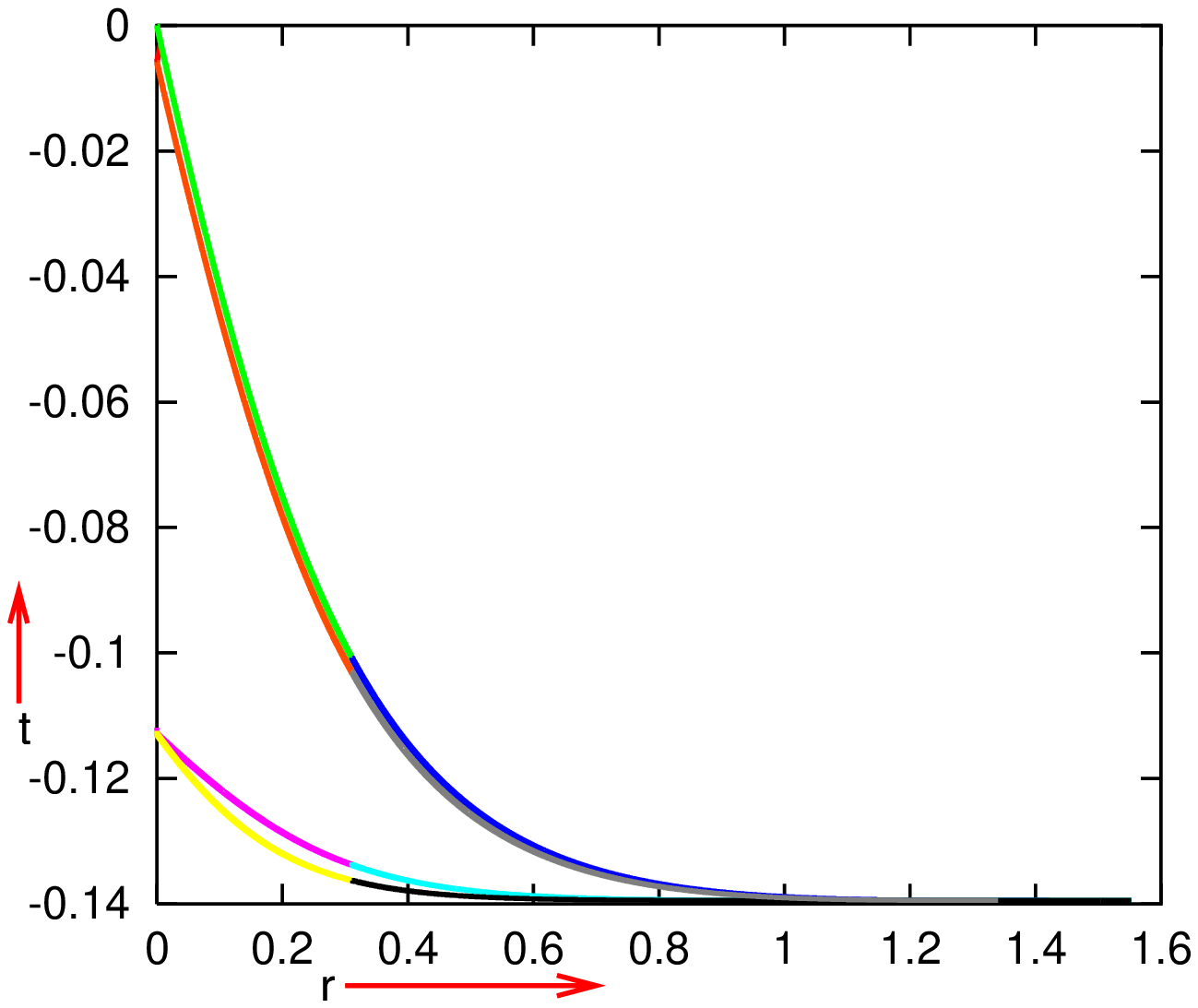}
${ }$ \\[-5.9cm]
\hspace{6.6cm}
\includegraphics[scale=0.45]{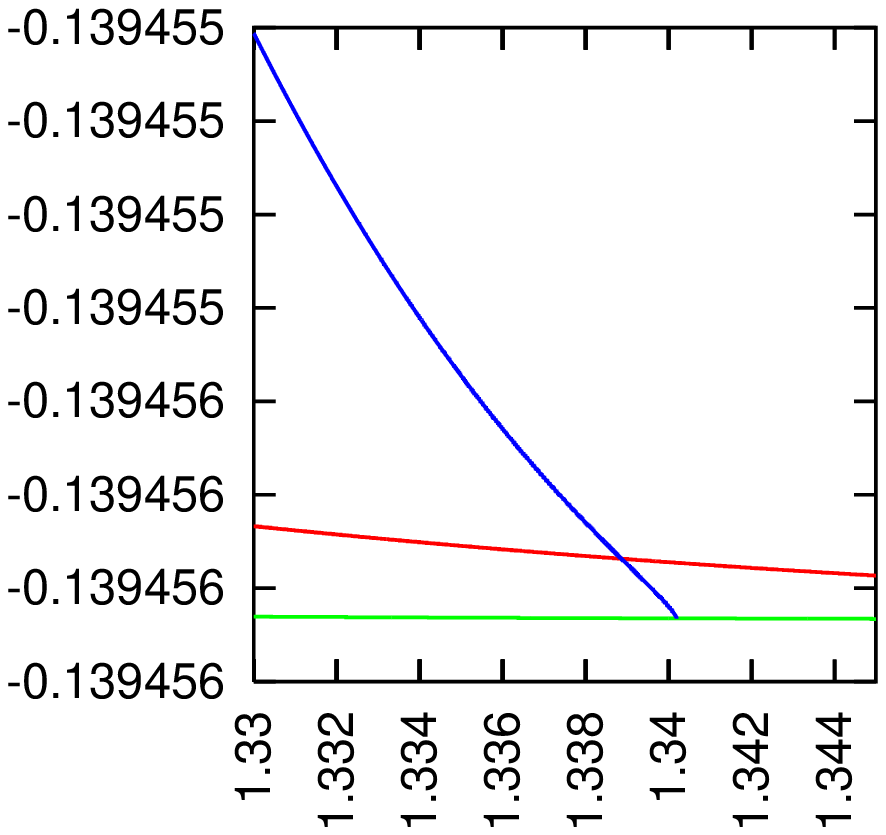}
\vspace{2cm}
\caption{{\bf Main panel:} The uppermost and the two lowest curves are those
from Fig. \ref{drawsecconeb}. The arc that nearly coincides with the cone
profile is ray N. See text for more explanation. {\bf Inset:} The neighbourhood,
where ray N hits the BB. The curves, counted from top to bottom at the left
edge, are ray N, $t_{\rm MRH}(r)$ and $t_B(r)$. The recombination time is
$\approx 3 \times 10^{-6}$ NTU above the upper edge of the graph. Redshift
never becomes negative along ray N, and becomes very large near the BB, see Fig.
\ref{drawsafez}. More explanation in the text. }
\label{drawsafecone}
\end{figure}

The third observer, O$_{\rm n}$ (for ``near''), is placed at such $r$ that the
ray she receives at the intersection with the PCPO is emitted from the BB where
the function $t_B$ is flat. The placement of O$_{\rm n}$ was determined by trial
and error. Its initial data at the PCPO are
\begin{eqnarray}
z_{\rm n} &=& 0.02000194389343255, \label{5.11} \\
r_{\rm n} &=& 0.00653692577372784, \label{5.12} \\
t_{\rm n} &=& -0.00293913865628162, \label{5.13} \\
R_{\rm n} &=& 0.002910104748843882. \label{5.14}
\end{eqnarray}
The ray emitted at the BB and received by O$_{\rm n}$ at the event given by
(\ref{5.12}) -- (\ref{5.13}) will be denoted N and is shown in Fig.
\ref{drawsafecone}. Similarly to ray OB, ray N, followed from the initial point
given by (\ref{5.12}) -- (\ref{5.13}) into the past, first reaches the center at
$t = -0.00581492733951897989$ NTU, then continues on the other side of the
center, hitting the BB at
\begin{equation}\label{5.15}
\left(\begin{array}{l}
r \\
t \\
\end{array}\right)_{\rm nBB} = \left(\begin{array}{l}
1.3401983891580524 \\
-0.13945554652960040 \\
\end{array}\right).
\end{equation}
At the contact of ray N with the BB, $t_B$ is constant up to better than $\Delta
t_B = 10^{-6}$ NTU $= 9.8 \times 10^4$ y.

\begin{figure}[h]
\begin{center}
\vspace{5mm}
\includegraphics[scale=0.54]{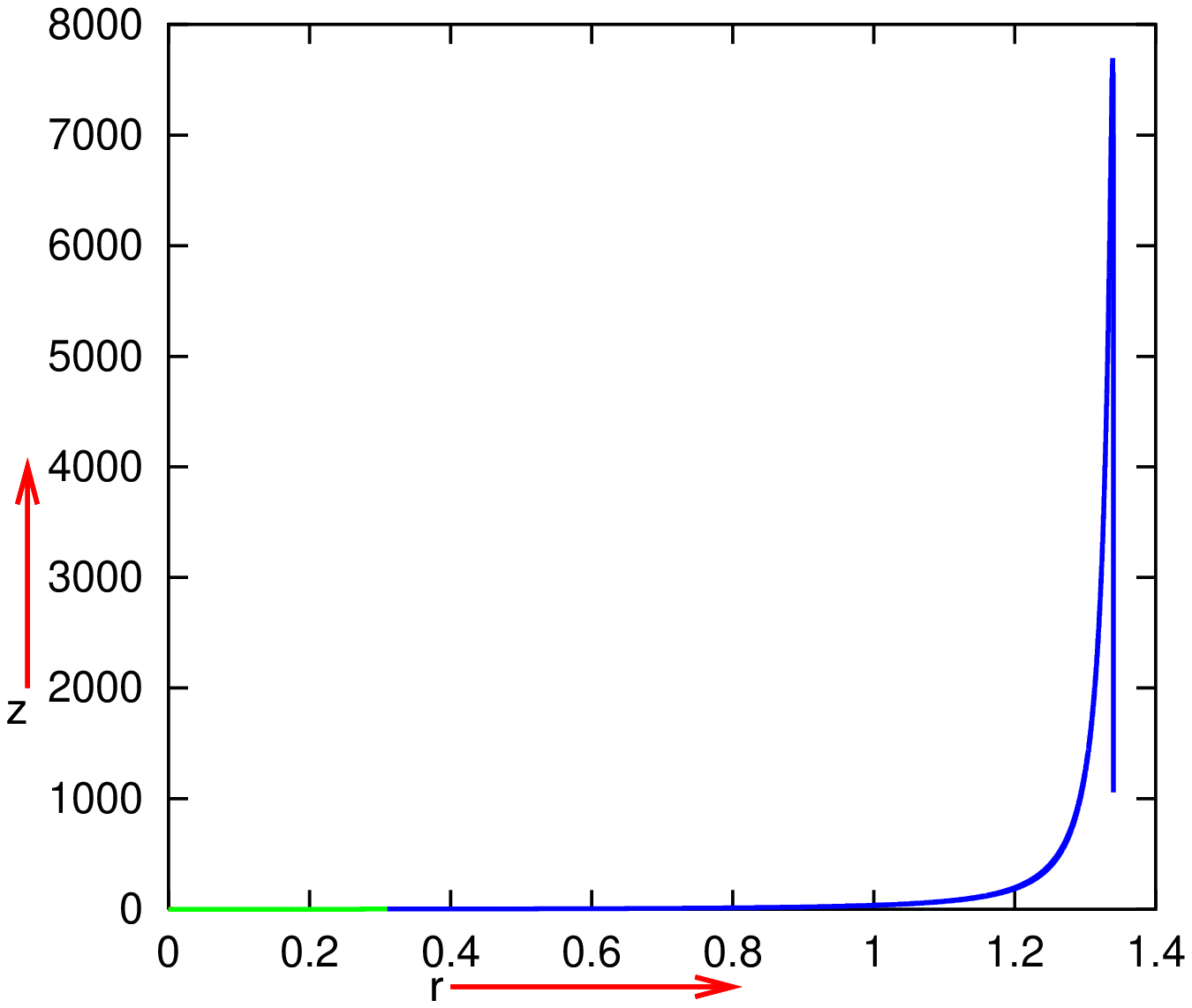}
${ }$ \\[-5.4cm]
\includegraphics[scale=0.4]{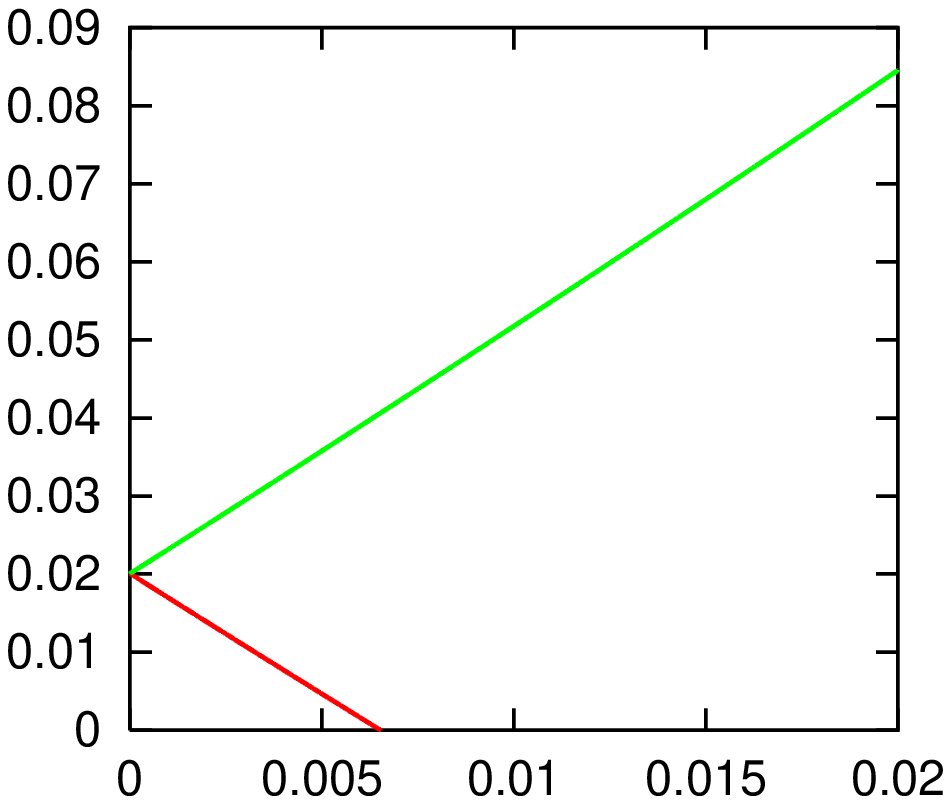}
\vspace{2cm}
\caption{{\bf Main panel:} Redshift $z(r)$ along ray N from Fig.
\ref{drawsafecone}. {\bf Inset:} Closeup view of the left end of the main graph.
See explanation in the text.}
\label{drawsafez}
\end{center}
\end{figure}

The redshift along ray N is shown in Fig. \ref{drawsafez}. The only point along
ray N where $z = 0$ is the initial point at the PCPO. Following ray N to the
past, $z$ attains the value 0.0200723615789212863 at $r = 0$, then increases up
to the maximum $z \approx 7676.412$, attained at
\begin{equation}\label{5.16}
\left(\begin{array}{l}
r \\
t \\
\end{array}\right)_{\rm nmax} = \left(\begin{array}{l}
1.3388618129278465 \\
-0.13945554652960040 \\
\end{array}\right).
\end{equation}
Then a numerical instability causes $z$ to go down at larger $r$. This is
because, at the level of precision assumed here, where ray N hits the BB, the
$t_B$ is ``not constant enough'' for $z$ to go to infinity. Were O$_{\rm n}$
placed nearer to the center, ray N would be indistinguishable from the PCPO.

\begin{figure}[h]
\hspace{-1.3cm}
${ }$ \\[1cm]
\includegraphics[scale=0.5]{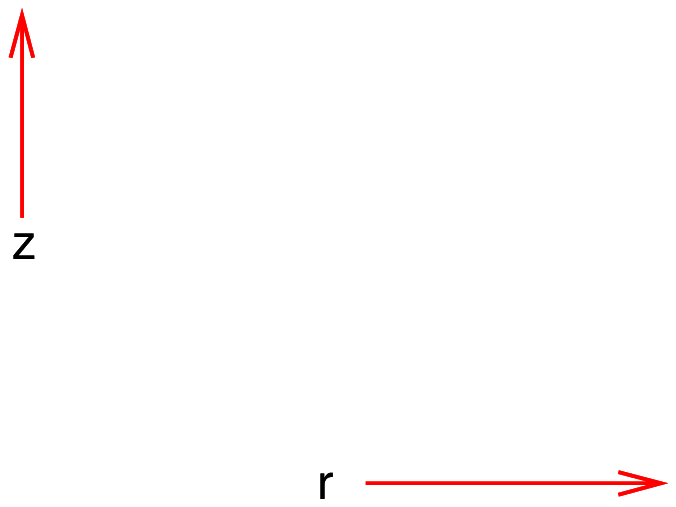}
${ }$ \\[-4.8cm]
\hspace{-3mm}
\includegraphics[scale=0.6]{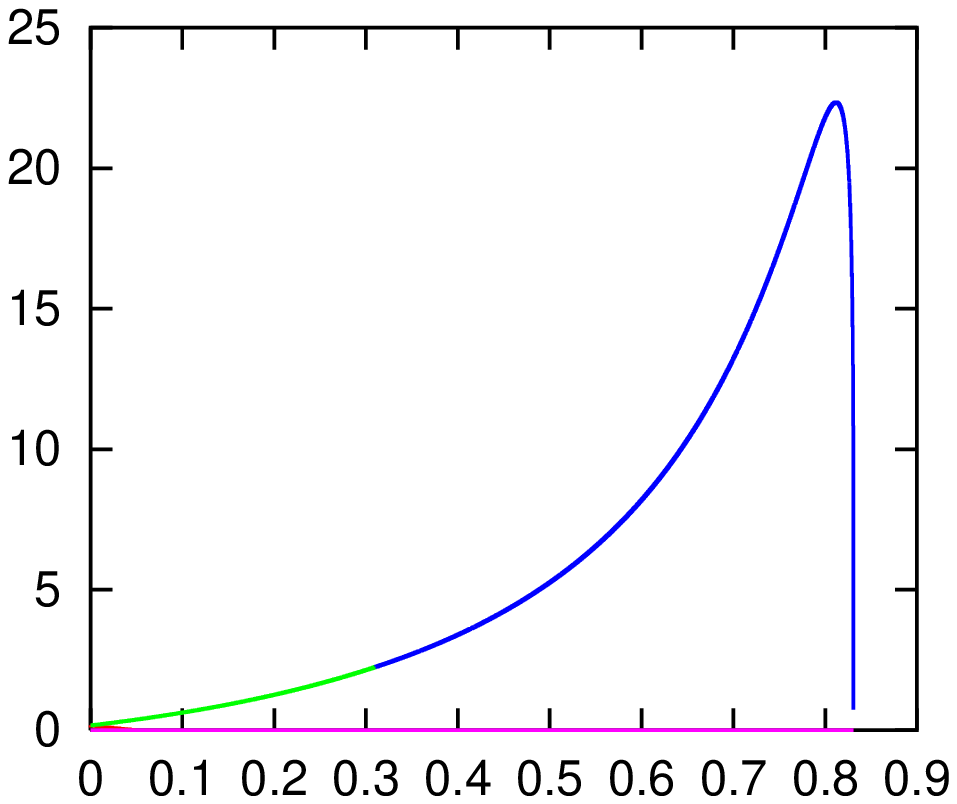}
${ }$ \\[-4.8cm]
\hspace{-3.5cm}
\includegraphics[scale=0.3]{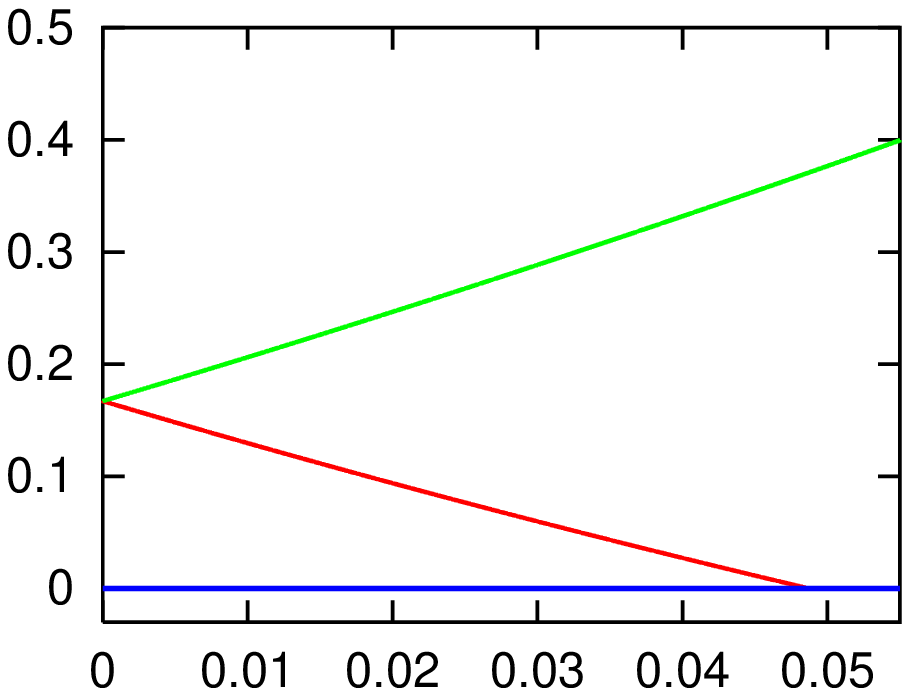}
\vspace{2.5cm}
\caption{Redshift $z(r)$ along a ray received by an observer placed between
O$_{\rm ob}$ and O$_{\rm n}$. See explanation in the text. The inset shows
$z(r)$ near $r = 0$. }
\label{drawzintermediate}
\end{figure}

The transition from rays emitted at nonconstant $t_B$ to those emitted at
constant $t_B$ is discontinuous. Beginning with the situation shown in Fig.
\ref{drawzstaresn} and moving the observer ever closer to the center, the
redshift profile changes as follows: the initial point where $z = 0$ moves
closer to the $r = 0$ axis, the point of crossing the $r = 0$ line moves down,
the maximum of $z(r)$ moves up and to the right, and the final segment of $z(r)$
that goes down becomes ever steeper, approaching vertical. If the observer is
close enough to the center, so that the ray (followed back in time) hits the BB
where $t_B$ is exactly constant, $z(r)$ goes to infinity at the contact with the
BB, and the final steep segment of the curve disappears. Figure
\ref{drawzintermediate} shows a graph of $z(r)$ intermediate between the
situations in Figs. \ref{drawzstaresn} and \ref{drawsafez}.

\section{The L--T($t_B$) model matched to Friedmann}\label{matching}

\setcounter{equation}{0}

Now we come back to the remark made in the paragraph after (\ref{5.10}). The
necessary and sufficient condition for matching an L--T to a Friedmann model
\cite{PlKr2006} is that at the boundary hypersurface the functions $M(r)$,
$E(r)$ and $t_B(r)$ go over in a continuous (not necessarily differentiable) way
into their Friedmann counterparts. Our functions $M(r)$ and $E(r)$ have
Friedmannian forms from the beginning. So, it is enough to assume that $t_B(r)$
becomes constant at the boundary value of $r$.\footnote{The model we consider
already coincides with Friedmann for $r > r_c$, where $r_c$ is given by
(\ref{4.6}). But to make the intended point, we need to match it to Friedmann at
$r_{\rm F}$ given by (\ref{6.1}).} As stated in the aforementioned remark, the
$r$ at the boundary should be between the $r_{\rm ob}$ given by (\ref{5.7}) and
$r = 0.2208$. So we choose the intermediate value $r_{\rm F}$, where
\begin{equation}\label{6.1}
\left(\begin{array}{l}
r \\
t_B \\
\end{array}\right)_{\rm F} = \left(\begin{array}{l}
0.2100014577175866 \\
-0.1325224690059549 \\
\end{array}\right).
\end{equation}
The time on the PCPO at this boundary is
\begin{equation}\label{6.2}
t_{\rm F} = -0.0778299400591163509\ {\rm NTU}.
\end{equation}
{}From the tables of values of $t(r)$ and $z(r)$ along ray OB we find that at
the $r_{\rm F}$ given above we have
\begin{equation}\label{6.3}
\left(\begin{array}{l}
t \\
z \\
\end{array}\right)_{\rm OB\ F} = \left(\begin{array}{l}
-0.13144220116062100 \\
2.2425612667236408 \\
\end{array}\right).
\end{equation}
Ray OB is now continued through $r = r_{\rm F}$ into the Friedmann region, with
the above as the initial data for it.

In the Friedmann region, (\ref{2.9}) simplifies to
\begin{equation}\label{6.4}
\dr t r = \pm \frac {S(t)} {\sqrt{1 - k r^2}},
\end{equation}
where $S(t) \df R/r$. Using (\ref{2.4}), (\ref{2.8}) and (\ref{2.17}), this can
be integrated with the result
\begin{equation}\label{6.5}
\eta(r) + C = \pm \ln \left(\sqrt{-k} r + \sqrt{1 - kr^2}\right),
\end{equation}
where $\eta(r)$ is the same as in (\ref{2.4}) and $C$ can be found from the
initial condition
\begin{equation}\label{6.6}
\eta_{\rm F} + C = \pm \left[\ln \left(\sqrt{-k} r + \sqrt{1 -
kr^2}\right)\right]_{\rm F},
\end{equation}
with $r_{\rm F}$ given by (\ref{6.1}), and $\eta_{\rm F}$ calculated from
(\ref{2.4}):
\begin{equation}\label{6.7}
\sinh \eta_{\rm F} - \eta_{\rm F} = \frac {(-k)^{3/2}} {M_0}\ \left(t_{\rm F} -
t_{\rm BF}\right).
\end{equation}
The $t_{\rm F}$ and $t_{\rm BF}$ are given by (\ref{6.1}) and (\ref{6.3}).

So, the construction of the Friedmann light cone and the calculation of redshift
along it goes as follows. The consecutive values of $r$ are taken from the same
table as in the previous calculations. Given the value of $r$, the $\eta(r)$ is
calculated from (\ref{6.5}) using (\ref{6.6}) for $C$, then $t(r)$ and $S(t(r))$
along the light cone are calculated from (\ref{2.4}) using (\ref{2.8}),
(\ref{2.17}) and (\ref{6.1}). Finally, with $S(t(r))$ known, the redshift along
the light cone is calculated from \cite{PlKr2006}
\begin{equation}\label{6.8}
1 + z(r) = z_{\rm F} + S_{\rm F}/S(t(r)).
\end{equation}

Figure \ref{drawFriedmanncone} shows the continuation of $t_B(r)$ (the lowest
line) and of ray OB through the boundary of the L--T and Friedmann
regions.\footnote{Ray OB, $t_B(r)$ and $z(r)$ can be made differentiable at $r =
r_{\rm F}$ by inserting an interpolating arc between the $t_B(r)$ of the L--T
region and the constant $t_B$ of the Friedmann region. The general matching
conditions do not require this \cite{PlKr2006}.} The line that ends at the
boundary is the MRH; it does not exist in the Friedmann region because there the
maximal redshift is infinite.

\begin{figure}
\includegraphics[scale=0.61]{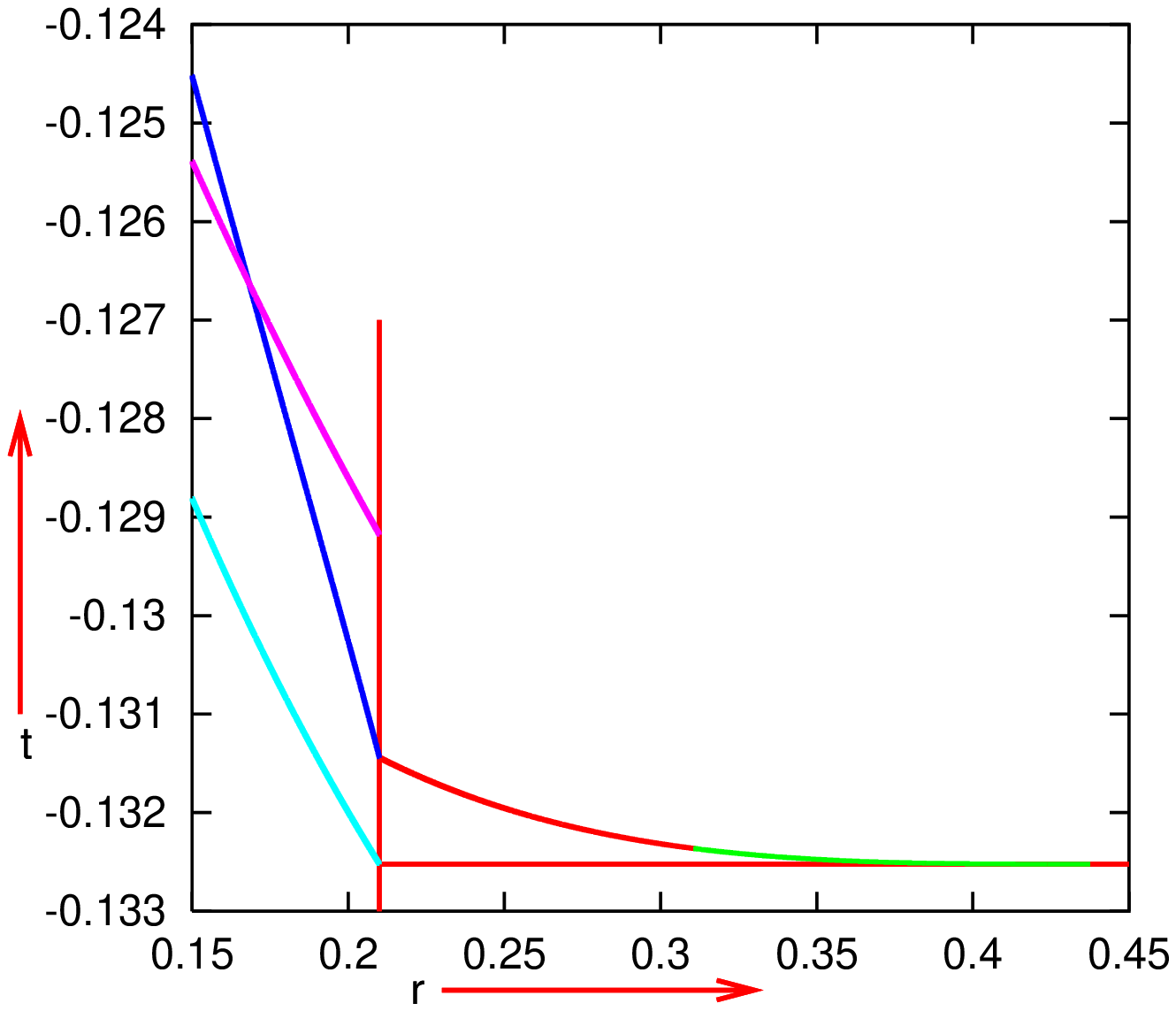}
${ }$ \\[-7.2cm]
\hspace{2.7cm}
\includegraphics[scale=0.45]{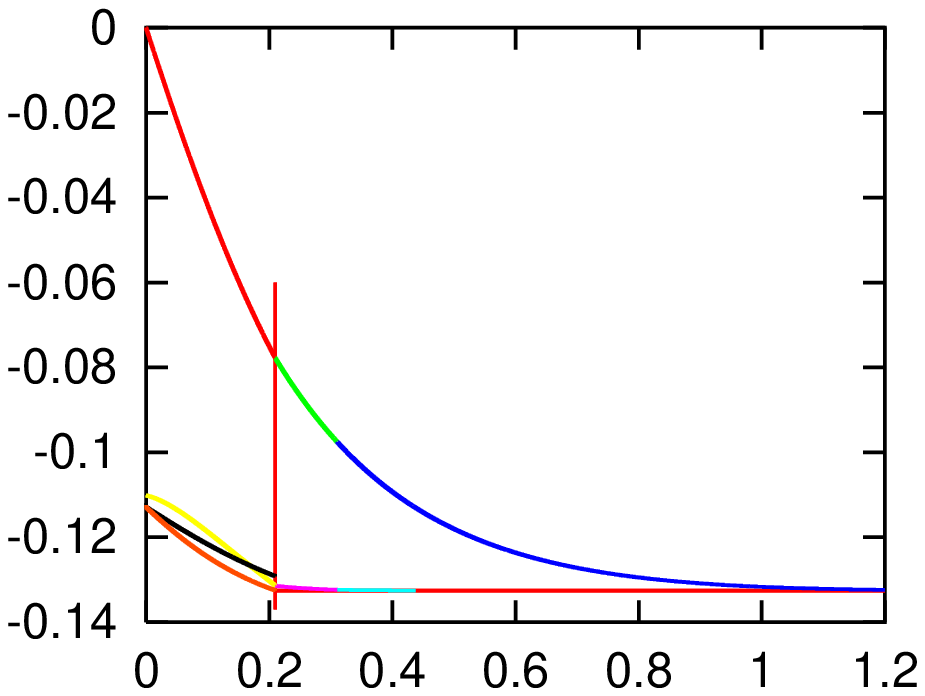}
\vspace{3.3cm}
\caption{{\bf Main panel:} Continuation of $t_B(r)$ (the lowest line) and of ray
OB into the Friedmann region. The vertical line marks the L--T/Friedmann
boundary at $r = r_{\rm F}$ given by (\ref{6.1}). The descending line that ends
at the boundary is the profile of the MRH. {\bf Inset:} The contents of the main
panel shown together with the complete past light cone of the present central
observer extended into the Friedmann region.}
\label{drawFriedmanncone}
\end{figure}

Figure \ref{drawFriedmannz} shows the continuation of $z(r)$ from Fig.
\ref{drawzstaresn} through the boundary of the L--T and Friedmann regions. The
function $z(r)$ was decreasing in the L--T region close to its boundary, but
becomes increasing in the Friedmann region, and increases until it becomes too
large to handle by the Fortran program. This happens at
\begin{equation}\label{6.9}
\left(\begin{array}{l}
r \\
z \\
\end{array}\right)_{\rm large} = \left(\begin{array}{l}
0.43753244000885227 \\
11861354545.253244 \\
\end{array}\right).
\end{equation}
This is, not accidentally, the same $r_{\rm large}$ at which the continuation of
ray OB into the Friedmann region becomes tangent to the constant-$t_B$ line.

\begin{figure}[h]
\hspace{-0.7cm}
\includegraphics[scale=0.63]{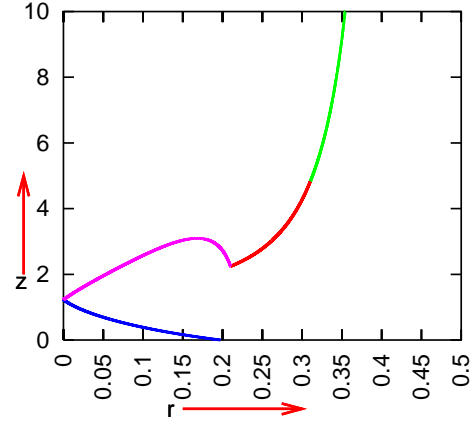}
\caption{Continuation of $z(r)$ from Fig. \ref{drawzstaresn} into the Friedmann
region. The redshift becomes too large to handle at $r = r_{\rm large}$ given by
(\ref{6.9}). } \label{drawFriedmannz}
\end{figure}

\begin{figure}[h]
\hspace{-0.4cm}
\includegraphics[scale=0.63]{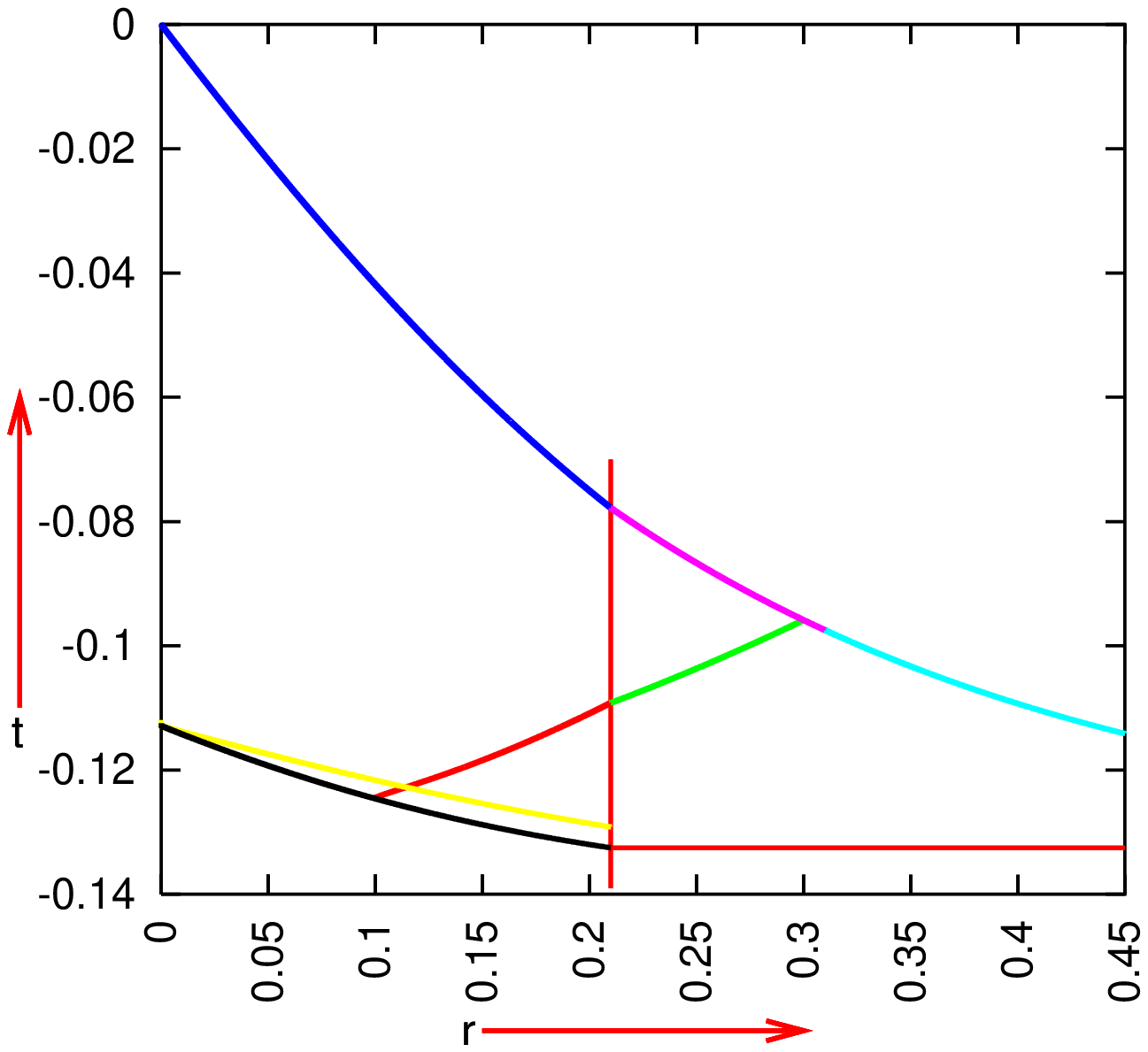}
${ }$ \\[-7cm]
\hspace{3.3cm}
\includegraphics[scale=0.4]{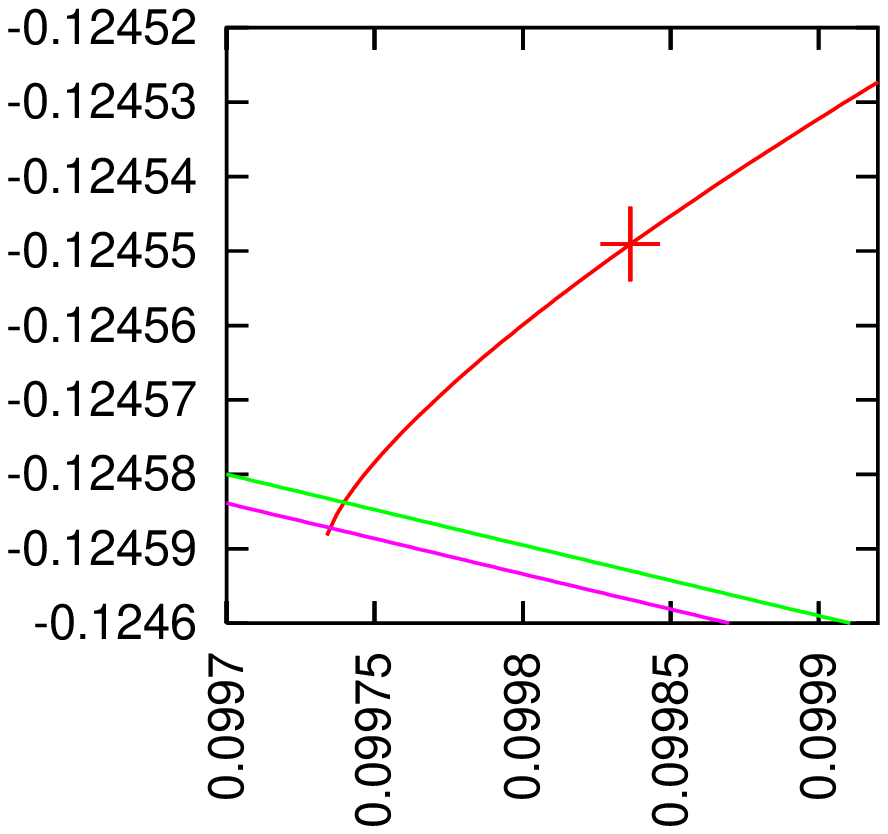}
\vspace{3.8cm}
\caption{{\bf Main panel:} The L--T($t_B$) model matched to Friedmann across $r
= r_{\rm F}$, and the ray crossing the boundary that displays blueshift in the
Friedmann region. See text for more explanation. {\bf Inset:} The
boundary-crossing ray in the neighbourhood of the Big Bang. The decreasing lines
are $t_B(r)$ (lower) and $t_{\rm rec}(r)$ given by (\ref{4.8}) (upper). The
cross marks the point on the ray where $z = 0$. The maximum-redshift
hypersurface is far above the upper margin.}
\label{drawblueFcone}
\end{figure}

The matching of the L--T and Friedmann models does not solve the problem of
blueshifts. The PCPO continued into the Friedmann region would still encounter
blueshifted rays emitted in the L--T($t_B$) region. Figure \ref{drawblueFcone}
shows one such exemplary ray. It was calculated in two stages:

1. Equations (\ref{6.5}) -- (\ref{6.8}) (with the $+$ sign in (\ref{6.5}) --
(\ref{6.6})) were used to calculate $t(r)$ and $z(r)$ back in time from the
initial point at the PCPO, with the coordinates
\begin{equation}\label{6.10}
\left(\begin{array}{l}
r \\
t \\
\end{array}\right)_{\rm ei} = \left(\begin{array}{l}
0.3000029697185931 \\
-0.0958721393954025947 \\
\end{array}\right).
\end{equation}
The ray reached the L--T/Friedmann boundary at
\begin{equation}\label{6.11}
\left(\begin{array}{l}
r \\
t \\
\end{array}\right)_{\rm eF} = \left(\begin{array}{l}
0.2100014577175866 \\
-0.10919095912654034 \\
\end{array}\right)
\end{equation}
with the redshift
\begin{equation}\label{6.12}
z_{\rm eF} = 0.37819933974218056.
\end{equation}

2. Using (\ref{6.11}) -- (\ref{6.12}) as initial data, (\ref{2.9}) --
(\ref{2.10}) (again with the $+$ sign) were integrated to determine the
continuation of $t(r)$ and $z(r)$ into the L--T region. The proximity of the
singularity did not allow $t(r)$ to end up with a vertical tangent, and the
$t(r)$ curve actually overshot the BB (as can be seen on close inspection of the
inset in Fig. \ref{drawblueFcone}). Its end point is at
\begin{equation}\label{6.13}
\left(\begin{array}{l}
r \\
t \\
\end{array}\right)_{\rm ee} = \left(\begin{array}{l}
0.0997339231053535613 \\
-0.12458819154593222 \\
\end{array}\right)
\end{equation}
with the last $z$ yet calculated being
\begin{equation}\label{6.14}
z_{\rm ee} = -0.8051202078031556.
\end{equation}

In addition to this ray, the main panel in Fig. \ref{drawblueFcone} shows the
PCPO (the uppermost curve) and $t_B(r)$ (the lowest curve), both continued into
the Friedmann region. The third decreasing line is the MRH profile, and the
vertical line marks the L--T/Friedmann boundary.

The inset in Fig. \ref{drawblueFcone} shows the final segment of the
boundary-crossing ray, on which $z(r)$ becomes negative. The coordinates of the
point, at which $z = 0$ are
\begin{equation}\label{6.15}
\left(\begin{array}{l}
r \\
t \\
\end{array}\right)_{{\rm e}z0} = \left(\begin{array}{l}
0.099836402226740395 \\
-0.12454907240930377 \\
\end{array}\right).
\end{equation}

\begin{figure}[h]
\hspace{-0.7cm}
\includegraphics[scale=0.63]{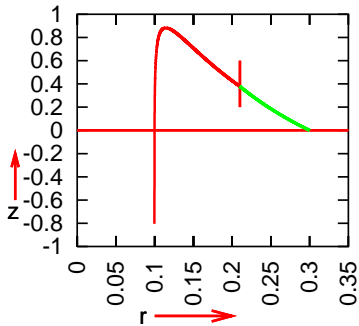}
\caption{The redshift along the boundary-crossing ray from Fig.
\ref{drawblueFcone}. The vertical bar marks $r = r_{\rm F}$ given by
(\ref{6.1}). } \label{drawblueFz}
\end{figure}

The profile of $z(r)$ along this ray is shown in Fig. \ref{drawblueFz}. The
coordinates of the maximum in $z$ are
\begin{equation}\label{6.16}
\left(\begin{array}{l}
r \\
t \\
\end{array}\right)_{\rm emax} = \left(\begin{array}{l}
0.11435427775654181 \\
-0.12276948639632553 \\
\end{array}\right)
\end{equation}
and the maximal value of $z$ is
\begin{equation}\label{6.17}
z_{\rm emax} = 0.88302700024316949.
\end{equation}
The point given by (\ref{6.16}) lies on the MRH profile up to better than
$\Delta t = 3.6 \times 10^{-8}$ NTU = 3528 y (the $r$ coordinates agree by
construction).

Comparing such a composite model with observations might be difficult. As seen
from Fig. \ref{drawFriedmannz}, the redshift along ray OB increases with $r$
only up to a maximum attained at $r_{\rm obmax}$ given by (\ref{5.10}), then
decreases with increasing $r$ up to the L--T/Friedmann boundary, and then starts
to increase again. In astronomy, it is assumed that redshift is a monotonically
increasing function of distance (and of look-back time); in fact, redshift is
routinely used as a measure of distance to objects far from the observer. To
test this model, a method of determining distance independent of redshift would
have to be introduced.

\section{Conclusions}\label{conclu}

\setcounter{equation}{0}

In a general L--T model with nonconstant $t_B(r)$, choosing $t_B$ nearly
constant (i.e., with a sufficiently small $\left|\dril {t_B} r\right|$), one can
move the maximum-redshift hypersurface (MRH) to times earlier than
recombination. At those times, the zero-pressure L--T model cannot describe the
Universe. Consequently, no blueshifts will be observed in the
after-recombination epoch (Sec.\ \ref{genLT}).

The rest of the paper is devoted to investigating blueshifts in one particular
L--T model, called L--T($t_B$). It is the one derived in Ref.\ \cite{Kras2014},
in which the $\Lambda$CDM function $D_L(z)$ is duplicated using nonconstant
$t_B$ alone; the $E(r) = - kr^2/2$ with $k$ given by (\ref{2.18}) is the same as
in a Friedmann model. In Sec.\ \ref{zmax}, the MRH is determined for this model.
In Sec.\ \ref{numcone}, the redshift/blueshift profiles along three
characteristic rays in this model are calculated and displayed. In Sec.\
\ref{matching}, the L--T($t_B$) model matched to Friedmann is investigated. The
matching hypersurface is chosen so that the L--T($t_B$) region encompasses all
the type Ia supernovae of the original project \cite{Ries1998,Perl1999}. This
matching does not solve the problem of blueshifts because observers in the
Friedmann region would receive blueshifted rays emitted from the nonconstant Big
Bang in the L--T($t_B$) region.

The final verdict on the L--T($t_B$) model is thus: if we insist on applying it
all the way back to the recombination time, then blueshifted rays will
inevitably cross the past light cone of the present central observer at
sufficiently large $z$. In the example shown in Fig. \ref{drawblueFcone},
blueshifts would be present beyond $z \approx 1.50087$. But this argument does
not ``rule out'' general L--T models with nonconstant $t_B$, as shown in Sec.\
\ref{genLT}.

To the attempts at discrediting the usefulness of the L--T model (or more
general ones) for cosmology, one can give a philosophical answer: objects
existing in Nature do not fulfil mathematical assumptions with perfect
precision. Assumptions such as spherical symmetry, axial symmetry, isolated
body, free fall, ideal gas, incompressible fluid, are in reality fulfilled only
up to some degree of approximation. Why should the Universe be an exception and
arise in an exactly simultaneous Big Bang, when the theory allows the BB to be
extended in (comoving) time?\footnote{By the way, why should it be exactly
homogeneous in the large and exactly spatially flat in addition?} Anticipating
more general solutions of Einstein's equations, one should even expect the most
general BB time to be a function of all three spatial variables, possibly
limited in generality by the constraint equations.

We generally agree that the Nature acts through mathematics. If so, then it is
reasonable to assume that it takes the tools from a generic set, e.g., not a
constant function when nonconstant ones are admissible, not a function of 2
variables when 3 are possible, etc. Would Nature ignore all this freedom in
order to keep the inflation hypothesis still alive and mainstream astronomers
feeling safe with their current knowledge?

\appendix

\section{Remarks about Ref.\ \cite{Zibi2011}}\label{critique}

\setcounter{equation}{0}

A comment on the terminology must be made here. The accelerated expansion of the
Universe is not an observed phenomenon. What is observed are redshifts and
apparent brightnesses of the SNIa. In the papers that first reported accelerated
expansion \cite{Ries1998,Perl1999}, these observations were interpreted using
exclusively the Friedmann models. Within this class of models, the best fit
between the model parameters and the observations is achieved when the curvature
index $k$ is zero, and the cosmological constant $\Lambda$ has a value that is
responsible for approx. 70\% of the present energy-density of the Universe (the
current figure is 68\% \cite{Plan2013}). The accelerated expansion, driven by
$\Lambda$, is thus a model-dependent element of interpretation of observations.

The earliest attempt at re-interpreting these observations using a less
simplistic model introduced a void around the center of symmetry
\cite{Tomi2000,Tomi2001a,Tomi2001b} (a lower-density Friedmann region surrounded
by a higher-density Friedmann background). This pioneering experiment caused
that the term ``void models of acceleration'' is now just reflexively used by
many authors to denote attempts at explaining the SNIa observations using models
with inhomogeneous matter distribution. This term is misleading and, in fact,
incorrect: several such models contain condensations instead of voids around the
center; see, for example, Refs.\ \cite{CBKr2010} and \cite{Kras2014b}.

The critical remarks presented below concern the concrete L--T model chosen by
the author as a basis for his considerations, but not the technical details of
Ref.\ \cite{Zibi2011}.

1. The author of Ref.\ [1] chose for his investigation the L--T model with
\begin{equation}\label{a.1}
E = 0 \qquad {\rm and} \qquad t_B(r) = a {\rm e}^{- (r/L)^2},
\end{equation}
where $a$ and $L$ are constants. The choice $E = 0$, justified by the desire to
consider only the ``decaying mode'' of perturbation of homogeneity, was too
special -- see points 2 and 4 below. The choice of $t_B(r)$ was not justified;
it was ``convenient''. Then, the author proved that the model defined by
(\ref{a.1}) does not pass the observational test of spectral distortions of the
light reaching the central observer, and used this as a basis for a far more
general claim that ``models with significant decaying mode contribution today
can be ruled out on the basis of the expected cosmic microwave background
spectral distortion''. This leap from the failure of one handpicked example
containing two arbitrary constants to the dismissal of the whole class labelled
by two arbitrary functions of $r$ is a logical error. By the same logic, one
could ``rule out'' the Robertson -- Walker models because one of them (the
Einstein Universe) is static, and so ``inconsistent with various observations''.
What Ref.\ \cite{Zibi2011} proved was only the fact that (\ref{a.1}) is not an
acceptable choice.

2. To consider the L--T model with a ``pure decaying mode'' in connection with
``void models of acceleration'', the function $t_B(r)$ must duplicate the
$D_L(z)$ of (\ref{2.14}) or another function fitted to the SNIa observations.
Thus, it cannot be freely chosen. In Ref.\ \cite{Kras2014} it was proven that,
assuming $E/M^{2/3} = - k$ constant (see point 4 below for justification), the
function $D_L(z)$ is duplicated only when $t_B(r)$ has a unique shape different
from (\ref{a.1}), while $k = - 4.7410812$. Consequently, the model (\ref{a.1})
is unrelated to the ``void models of acceleration''.

3. The ``decaying mode'' was chosen for a kill because the author was convinced
that the ``pure growing mode'' had been ruled out already before. Isolating the
growing and decaying modes for separate investigations is not productive -- see
the remarks in Ref.\ \cite{Kras2014}. As shown in Sec.\ \ref{genLT} here, a
combination of the two modes allows us to avoid the problem with blueshifts
completely.\footnote{As proved in Ref.\ \cite{Kras2014}, when $E = 0$, the
$D_L(z)$ relation of the $\Lambda$CDM model cannot be reproduced. This case was
included in Sec.\ \ref{genLT} for completeness only.} The vocabulary of growing
and decaying modes is borrowed from investigations in linearized Einstein
equations and is not useful in the exact theory. Within the exact theory, it is
more instructive to deal with quantities that have direct physical meaning, for
example, with density or velocity distributions at different hypersurfaces of
constant time \cite{KrHe2004,KrHe2004b,BKHe2005}.

4. Even so, the ``pure decaying mode'' is not correctly defined in Ref.\
\cite{Zibi2011}. The increasing perturbations of homogeneity are not generated
by ``small-amplitude variations in spatial curvature'', but by variations, not
necessarily small, in the function $E / M^{2/3}$. Thus, to isolate a pure
decaying mode, one must assume $E / M^{2/3} =$ constant, as was done in Refs.\
\cite{Kras2014,INNa2002,YKNa2008} and \cite{Yoo2010}, and not $E = 0$, as did
the author of Ref.\ \cite{Zibi2011}.

5. With $t_B(r)$ given by (\ref{a.1}), the function $\dril {t_B} r$ is nonzero
\textit{for all $r$} (unlike in Ref.\ \cite{Kras2014} and in the present paper,
where $\dril {t_B} r$ becomes exactly zero at a finite $r$). Therefore, even the
primary CMB ray, along which spectral distortions are calculated in Ref.\
\cite{Zibi2011}, will reach the present central observer with a blueshift,
unless the values of $a$ and $L$ are suitably chosen. The intersection of this
ray with the ZRH can be pushed to before the recombination epoch by choosing $L$
sufficiently small, but Ref.\ \cite{Zibi2011} does not say whether this was
actually taken care of. Without this, one considers blueshifts being distorted
by blueshifts.

\section{The proof of Lemma 3.1}\label{provelem3.1}

\setcounter{equation} {0}

Let $- t_{B,r} = \varepsilon > 0$ at a given $r$ (from (\ref{3.5})). We shall
prove that $t \to t_B$ when $\varepsilon \to 0$.

In constructing the proof it must be remembered that, given $t_B(r)$, the
function $E(r)$ is determined by (\ref{2.13}) -- (\ref{2.14}). Consequently,
when $\left|t_{B,r}\right|$ is decreased at a given $r$, we have to take into
account that $E,_r/(2E)$ in (\ref{3.3}) will thereby also change, and we do not
know how.

Going from $- t_{B,r} = \varepsilon > 0$ to a lower value $- t_{B,r} =
\varepsilon_2 > 0$ causes the product on the left-hand side of (\ref{3.3}) to
also decrease. If it is $(t - t_B)$ that decreases, then the thesis is proved.
If it is the other factor that decreases, then, in a region with no shell
crossings, it may not become smaller than $3/r$. This is because $F_1(\eta) > 0$
(from (\ref{3.6})), $E > 0$ (by assumption in this case) and $E,_r > 0$ (when
there are no shell crossings), while $3/r$ is independent of $t_{B,r}$: the
value of $r$ only depends on the point at which the whole analysis is done.
Since we can make $(- t_{B,r})$ arbitrarily small, and $\left[\frac {E,_r} {2E}\
F_1(\eta) + \frac 3 r\right]$ has the lower bound $3/r$, decreasing $(-
t_{B,r})$ will eventually cause $(t - t_B)$ to decrease, too. $\square$

This proof fails when shell crossings are present $(E,_r < 0)$ -- see the
footnote in the paragraph below (\ref{3.5}).

\bigskip

{\bf Acknowledgements:} I am grateful to Micha{\l} Szurek for his help in
formulating the concluding paragraph, and to Jim Zibin for a discussion that
helped to clarify a few points.

\end{document}